\documentclass[%
 reprint,
 amsmath,amssymb,
 aps,
]{revtex4-2}

\usepackage{graphicx}
\usepackage{dcolumn}
\usepackage{bm}
\usepackage{xcolor}
\usepackage{hyperref}
\usepackage{url}

\begin{document}

\preprint{APS/123-QED}

\title{Spiking Rate and Latency Encoding with Resonant Tunnelling Diode Neuron Circuits and Design Influences} 

\author{Giovanni Donati$^1$}
\author{Dafydd Owen-Newns$^1$}
\author{Joshua Robertson$^1$}
\author{Xavier Porte$^1$}
\author{Ekaterina Malysheva$^2$}
\author{Jose Figueiredo$^3$}
\author{Bruno Romeira$^4$}
\author{Victor Dolores-Calzadilla$^2$}
\author{Antonio Hurtado$^1$}%
 \affiliation{ $^1$Institute of Photonics, SUPA Dept of Physics, University of Strathclyde, Glasgow, United Kingdom}%
 \affiliation{$^2$Eindhoven Hendrik Casimir Institute, Eindhoven University of Technology, Eindhoven, Netherlands}%
 \affiliation{$^3$LIP and Departamento de Física de Faculdade de Ciências de Universidade de Lisboa, Lisboa, Portugal}%
 \affiliation{$^4$International Iberian Nanotechnology Lab., Braga, Portugal}

\date{\today}

\begin{abstract}
Neuromorphic computing, inspired by the functionality and efficiency of biological neural systems, holds promise for advancing artificial intelligence and computational paradigms. Resonant tunneling diodes (RTDs), thanks to their ability to generate neuronal dynamical responses, such as excitable spiking and refractoriness, have recently emerged as candidates for use as opto-electronic spiking neurons in novel neuromorphic computing hardware. This work explores the ability of RTD spiking neurons to deliver information encoding mechanisms analogous to those observed in biological neurons, specifically spike firing rate and spike latency encoding. We also report detailed experimental and numerical studies on the impact that the RTD mesa radius and circuit inductance and capacitance have on its spiking properties, providing useful information for future design of RTD-based neural networks optimised for ultrafast ($>$GHz rate) processing capabilities. Finally, we showcase the application of spike rate encoding in RTD neurons for the ultrafast reconstruction of a eight-level amplitude signal, effectively filtering out the noise.
\end{abstract}

\maketitle


\section{Introduction}
Within the artificial intelligence landscape, neuromorphic systems are emerging as ensembles of nonlinear nodes that mimic the way biological brains process information. While artificial neural networks (ANNs) were initially emulated serially on standard CPUs—leading to significant breakthroughs in supervised and unsupervised learning \cite{schuman2022opportunities,eshraghian2023training}—the slowing of Moore’s Law \cite{waldrop2016chips} has emphasized the need for computing platforms that mirror the inherent parallelism occurring in the brain. Both electronic and photonic platforms are under active development, integrating trainable physical parameters and neuron-like units. Examples of neuromorphic electronic hardware include SpiNNaker from the University of Manchester \cite{furber2014spinnaker,painkras2013spinnaker}, IBM’s TrueNorth \cite{debole2019truenorth}, Intel’s Loihi \cite{davies2018loihi}, and BrainScaleS from the University of Heidelberg \cite{schemmel2017accelerated}. Neuromorphic photonics, on the other hand, leverages the high speed, low loss, and parallel processing enabled by light, employing approaches such as wavelength division multiplexing in both on-chip \cite{ashtiani2022chip, biasi2023array, feldmann2019all} and free-space configurations \cite{lin2018all}.

Resonant tunneling diodes (RTD) have recently gained attention for their operational versatility as opto-electronic spiking neurons.
These active semiconductor devices incorporate a double-barrier quantum well (DBQW) structure with layer thicknesses of approximately 10 nm. This preserves the electronic wave function’s phase across the DBQW enabling interference phenomena at room temperature \cite{ferry2020transport}. 
RTDs operate analogously to optical (tunneling in) tunable fabry-perot filter/interferometer: Applying a voltage across the DBQW adjusts the Fermi energy of incident electrons maximizing electron flow tunneling when their energy resonates with the DBQW states. 
This results in RTD's N-shape current-voltage (I-V) characteristics with a positive differential resistance (PDR) region from zero volt up to the resonance point (at peak voltage, where the current reaches a local maximum). Beyond this, tunneling is inhibited, leading to a negative differential resistance (NDR) region up to the valley voltage where the current reaches a minimum \cite{ironside2023resonant}. 
The resulting highly nonlinear I-V characteristics of RTDs has been used to demonstrate ultra-high (terahertz-range) bandwidth switching capabilities between the peak and valley of the I-V curve, making them attractive for communication applications \cite{cimbri2022resonant, nishida2019terahertz}, and also diverse neuromorphic behaviours. These include excitable spike firing \cite{romeira2013excitability, hejda2022resonant}, refractoriness \cite{hejda2023artificial}, multi-pulse bursting \cite{ortega2021bursting}, oscillatory and chaotic responses \cite{romeira2014stochastic}, flip-flop memory oscillating between a spiking and quiescent states \cite{donati2024spiking} or two stable values \cite{sano200180}, and regenerative self-feedback operation \cite{romeira2017delay}. These behaviours can be exploited in isolated RTDs or when coupled with external optoelectronic elements, such as RTD-laser diode systems integrated on-chip \cite{slight2007investigation, romeira2013excitability}. Despite progress, the encoding mechanisms used by RTD-based neurons and their parallels with their biological counterparts remain underexplored.

This study experimentally explores for the first time spike rate and spike latency encoding mechanisms (at fast nanosecond rates) in RTD neuron circuits, further aligning their functionality with biological neurons. 
We also investigate experimentally and in theory the impact of design parameters, in particular the RTD cross-sectional area and circuit inductance and capacitance, on their neuromorphic rate-encoding properties and potentials for ultrafast speed operation at $>$GHz rates. These findings will be valuable for guiding the design and training of future neuromorphic hardware based on opto-electronic spiking neurons built with RTDs. Finally, we propose a proof-of-concept application where a 8 level noisy voltage signal is regenerated by an RTD neuron by means of its rate encoding capability.

The paper is structured as follows: Section \ref{sec_IE} presents experimental results on the rate and latency encoding capabilities of a single ($\mu$m-scale) RTD spiking neuron. Section \ref{sec_design} explores the effects that mesa radius, inductance and capacitance, key design parameters, have on the neuromorphic properties and speed operation of RTD neurons. Then, in Section \ref{sec_app} we propose a proof-of-concept application for encoding, reconstruction and noise-filtering of a multi-level signal leveraging the ultrafast spike rate encoding capabilities of RTD neurons. Finally, in section \ref{sec_conclusions} we draw the conclusions of the study.

\section{Information encoding}
\label{sec_IE}

\begin{figure}[t]
\includegraphics[width=0.5\textwidth]{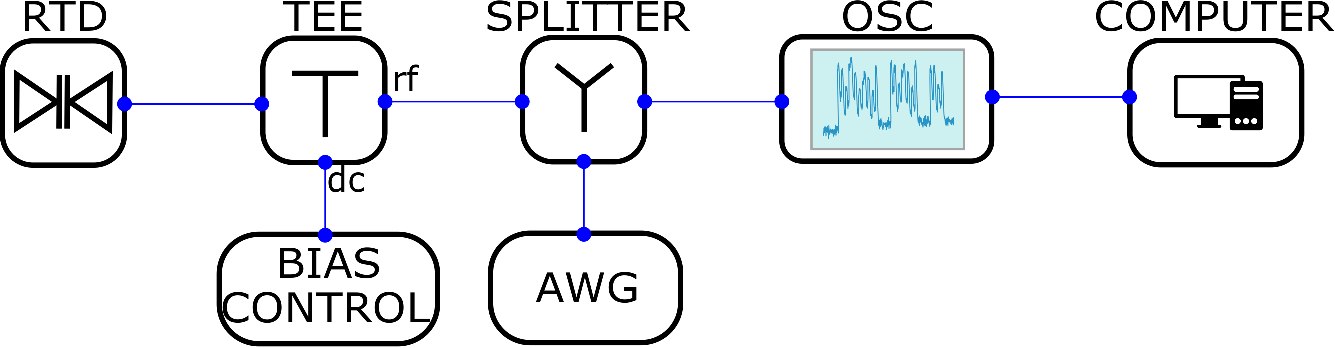}
\caption{Experimental setup for probing the RTD. AWG: arbitrary waveform generator, OSC: oscilloscope, TEE: Bias tee.}
\label{fig_exp_scheme}
\end{figure}
We used the setup sketched in Fig. \ref{fig_exp_scheme} to experimentally investigate the information econding capabilities of RTD neurons. The IV characteristic of RTDs is measured through a bias controller (keysight E36312A), which both sets the desired bias voltage to the RTD and reads the corresponding output current. An arbitrary waveform generator (AWG, keysight M8190A, 12 GSa/s) allows generating rf waveforms with voltage pulses that will be used to elicit spiking dynamics in the RTD neuron. These pulses propagate towards a splitter and then a bias tee ([0.01-12.4] GHz bandwidth range), where they are combined with the bias dc voltage signal. The temporal dynamical evolution in the system at its fast (spiking) timescales is measured by directing the rf component of the RTD output to a high-speed oscilloscope (OSC, Rohde $\&$ Schwarz, 16 GHz bandwidth) which is then connected to a computer for data storage and analysis.
All the RTDs of this work were fabricated on a layerstack grown by metal-organic vapor phase epitaxy (MOVPE) on a semi-insulating InP substrate. They contain a 1.7 nm AlAs/5.7 nm InGaAs/1.7 nm AlAs DBQW structure surrounded by 2 nm undoped InGaAs layers, followed by 50 nm of moderately doped n-InGaAs ($5\times10^{16}$ cm$^{-3}$) and highly doped n-InGaAs contact layers.

\subsection{Spike rate encoding}

\begin{figure}[t]
\includegraphics[width=0.5\textwidth]{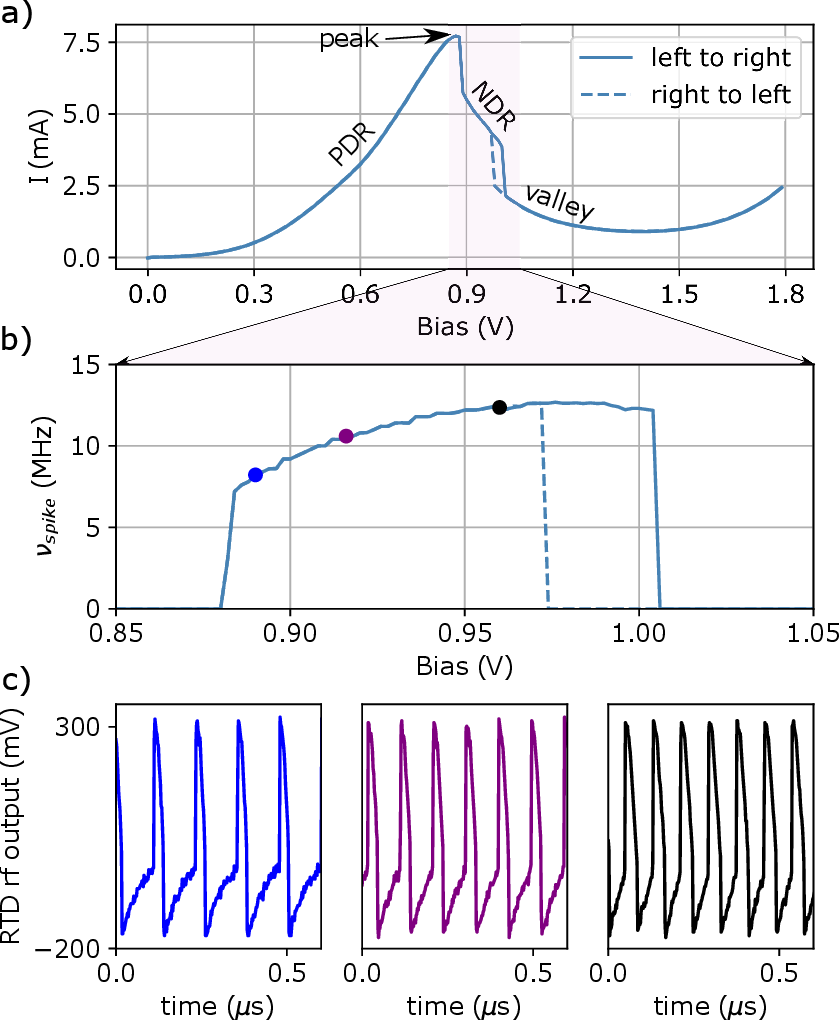}
\caption{a) Experimental I-V characteristics of an RTD with a 3 $\mu m$ circular mesa radius. b) Spiking rate variation under dc bias tuned across the NDR region. c) Examples of spike time traces with increasing firing rate achieved for three bias values within the NDR region, labeled by relative color circle in b).}
\label{fig_IV_3um}
\end{figure}

The dynamical behavior of an RTD neuron circuit evolves as the bias voltage is moved within the NDR (Negative Differential Resistance) region of its I-V characteristic. Referring to the experimentally-measured I-V curve (see Fig. \ref{fig_IV_3um}(a)) of a single-pillar RTD with a 3 $\mu$m mesa radius, the system remains quiescent when biased below the peak ($V_{peak}=0.88$ V) voltage of the I-V curve. However, surpassing the peak voltage initiates oscillatory dynamics, which persist until the bias approaches the valley region, beyond which the system becomes quiescent again.
When biased within the NDR region, the RTD circuit exhibits self-oscillations with a spike-like profile. This behavior is captured in Fig. \ref{fig_IV_3um}(b), where the spiking firing rate of the RTD, $\nu_{spike}$, increases with the bias voltage, nearly doubling from from 7 MHz to 12.5 MHz. Additionally, Fig. \ref{fig_IV_3um}(c) presents representative spiking time traces for three distinct bias scenarios, corresponding to coloured markers in Fig. \ref{fig_IV_3um}(b). These traces illustrate varying numbers of spikes generated over 0.5 $\mu$s, reflecting the correlation between bias voltage and firing rate.
Interestingly, the firing rate behavior of the RTD neuron closely resembles that of a class II biological neuron, suggesting a biologically inspired mechanism for encoding information in the firing rate. To utilize this simple yet effective encoding method, the RTD can be biased just below the peak voltage in its I-V characteristic, where it remains quiescent. Input voltage perturbations can then temporarily drive the RTD into the NDR region, enabling spiking. Input pulses of varying intensity modulate the bias within the NDR region, producing spike trains whose firing rates depend on the amplitude of the input pulse. Alternatively, the bias itself can be varied across the NDR region to produce the same spike-rate variations. 

For effective information encoding, the duration of the perturbation must be carefully managed. Voltage pulses that are too short may trigger only a single spike (not enough to define a firing-rate), while excessively long pulses may generate unnecessary long spike-trains, reducing processing speed. An optimal compromise should be achieved by considering the timescales of the task of interest, and therefore the speed at which information needs to be converted and processed. 
When the RTD is biased close but below the peak value (in its I-V curve), the amplitude of the input voltage pulse required to drive the RTD into the NDR region depends on the applied bias voltage. Consequently, the bias emerges as a trainable parameter during the learning phase of a neuromorphic task. 
Notably, the three time-series examples in Fig. \ref{fig_IV_3um}(c) also highlight that while the spike firing rate varies for different bias values within the NDR region, the amplitude of the spike remains constant.

\begin{figure}[t]
\includegraphics[width=0.5\textwidth]{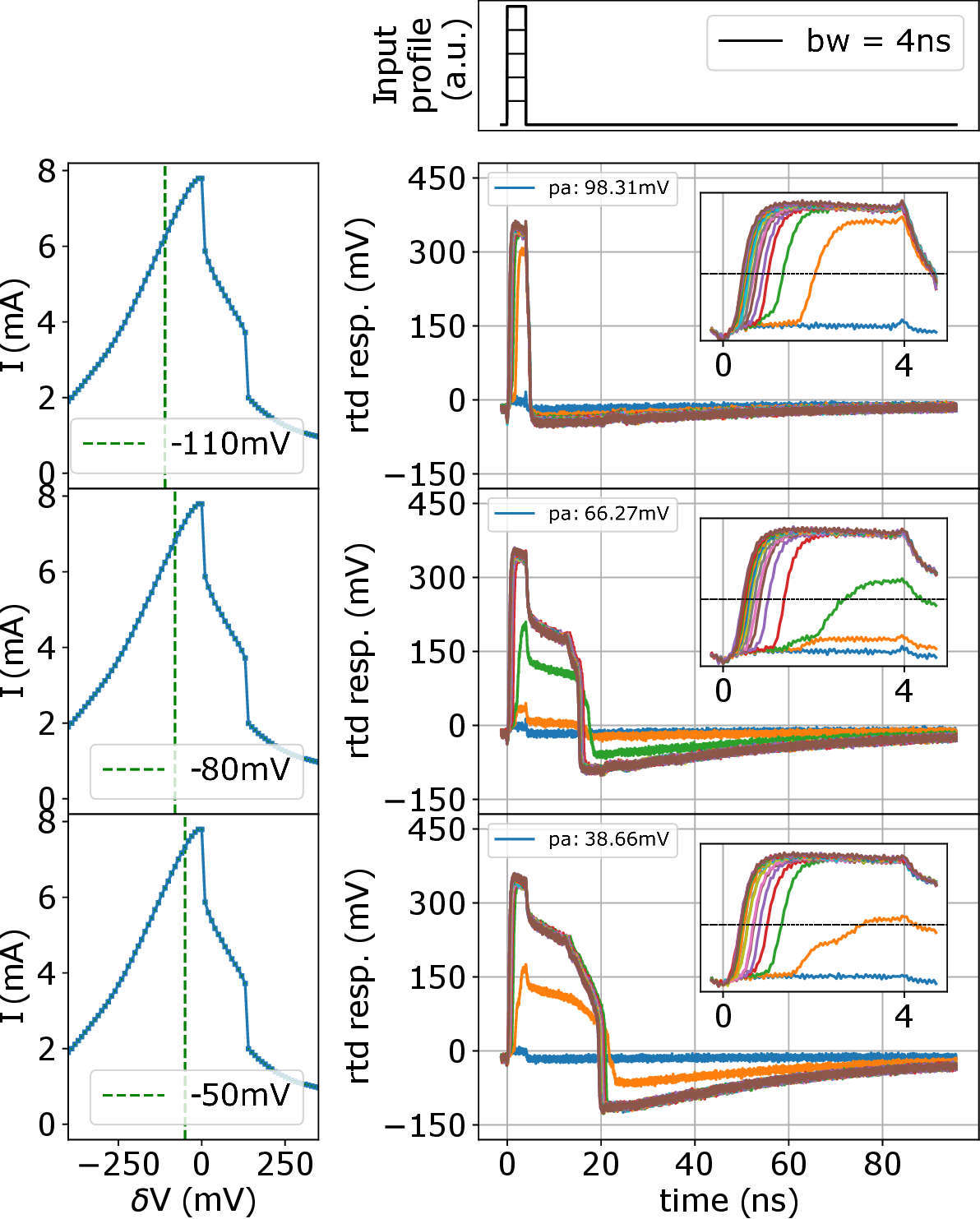}
\caption{Examples of experimental RTD rf responses for three bias scenarios (green dashed line in first column panels) far from the peak ($V_{peak}=0.88$ V), and input pulses lasting 4 ns and having increasing amplitudes (top panel). Blue curves in centered panels describe the quiescent RTD response to input pulses with amplitude just below the spike activation threshold. Higher amplitude pulses (with 1.25 mV steps) evoke the RTD spiking response, with a latency that reduces as sketched in the insets. }
\label{fig_time_encoding}
\end{figure}

\begin{figure}[t]
\includegraphics[width=0.4\textwidth]{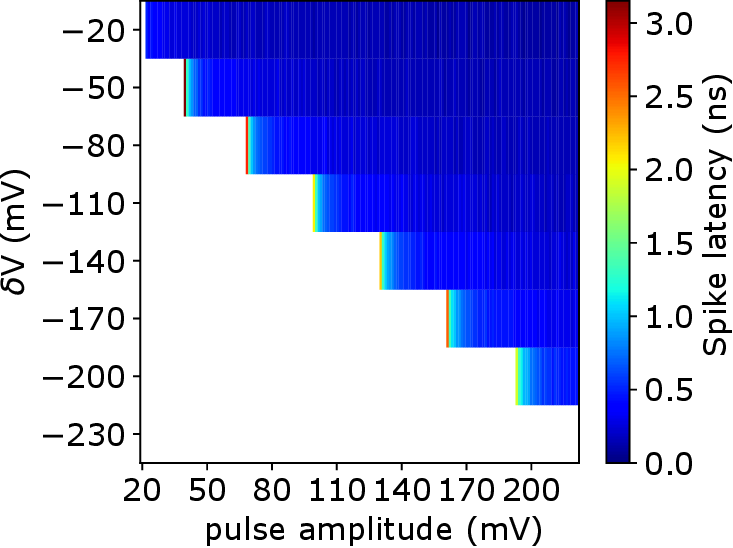}
\caption{Spike latency map for different configurations of bias $\delta V = V_{bias}-V_{peak}$ (y axis) and input pulse amplitude (x-axis). The input pulse has a duration of 4 ns. When its amplitude is not sufficiently strong to elicit an RTD spike response, the delay is not defined, appearing as a white box. }
\label{fig_time_summary}
\end{figure}

\subsection{Spike latency encoding}
Information can also be encoded in the precise timing of a neuron's spike, offering a faster alternative to rate encoding. In this scheme, a single spike can encode information, eliminating the need to generate multiple spikes to define a rate. Specifically, we examined the relationship between the amplitude of an input perturbation and the latency time of the spike it triggers in an RTD neuron circuit. The latter, known as latency encoding, is a powerful mechanism widely used by the brain to extract spatial features from somatosensory, visual and auditory sensing information \cite{richmond1987temporal, middlebrooks1994panoramic}. 

Latency encoding is investigated with the same RTD neuron previously used for rate encoding, having a 3 $\mu$m mesa-radius. 
Here, we bias the RTD below the peak (of its I-V curve), thus preparing it in a quiescent state. Then, we perturb the RTD with voltage pulses of varying amplitudes but fixed duration, much shorter than the duration of a single spike (and that of its refractory period, approximately 90 ns), to elicit a single spike. 
Examples of RTD dynamics are shown in Fig. \ref{fig_time_encoding}. The selected bias is indicated as green dashed lines in the I-V curves in the first column and expressed in terms of $\delta V = V_{bias}-V_{peak}$, i.e. its distance from the peak of the I-V curve ($V_{peak}=0.88$ V). The temporal profiles of increasing amplitude input pulses with 4 ns duration are shown in the first top panel.
Input pulses with amplitude just below the threshold do not trigger any spike (blue time traces in centered panels). Sup-threshold pulses (with resolution of 1.25 mV) enable the spike response with a latency given by the time taken to intersect an arbitrary voltage threshold represented as horizontal black line. As the input strength increases, the spike is generated fist (lower latency).
A wider characterization of the spike's latency for input pulses lasting 4 ns and increasing amplitude from 20 mV to 251.25 mV with 1.25 mV steps, and different bias conditions $\delta V$ from -20 mV to -230 mV with 30 mV step, is summarized in Fig. \ref{fig_time_summary}. The pulse amplitude used to evoke the RTD spiking response is reported on the x-axis, with 1.25 mV steps resolution. The resulting latency at which spikes are generated are presented in a colourmap, with white regions indicating quiescent states where no spike is fired by the RTD circuit. 
The results reveal that the transition from quiescence to spiking RTD dynamics is strongly dependent on both the voltage bias applied to the RTD and the amplitude of the input pulse. For any tested bias, spikes are elicited only when the pulse amplitude approaches or exceeds $\delta V$. 
The closer the bias is to the peak, i.e. $\delta V \rightarrow$ 0, and the lower is the perturbation amplitude needed to bring the device into the NDR region of the I-V curve, and elicit a spike. 
The latency of the spike is inversely proportional to the input pulse amplitude for a few mV-range just above the threshold voltage. Spikes are fired with the longest delay for input amplitudes just above the threshold, and earlier for larger amplitudes, rapidly reaching a plateau in all bias $\delta V$ considered. Moreover, this behavior is consistently reproduced in all other measures with different pulse durations investigated (2 ns, 1 ns, and 0.5 ns, not shown), indicating a mechanism where the timing of the spike depends on both a restricted range of pulse amplitude values (of just a few mV) above threshold and on the operating bias. Given this reduced input voltage range, we conclude this encoding mechanism working only for differentiation of closely time-resolved input pulses. 
We will show in the following that tuning just the circuitry inductance allows for sub-ns spikes while leaving almost unchanged the integration time. Within this scenario, latency up to 3-ns can play a major role, enabling time-to-first-spike protocols. 
For example, latency encoding is used to extract spatial information from sensory data in biological neurons \cite{richmond1987temporal, middlebrooks1994panoramic}. Similarly, an array of RTDs firing each one with different latencies may be studied in future designs. Despite the sensing voltage range would be restricted to only few mV, according to Fig. \ref{fig_time_summary}, this range could be widened by designing a matrix of RTDs, having different bias lines per row. In this way, each row of RTDs would be highly sensitive in a restricted but different voltage range.

\begin{figure*}[t]
\includegraphics[width=1\textwidth]{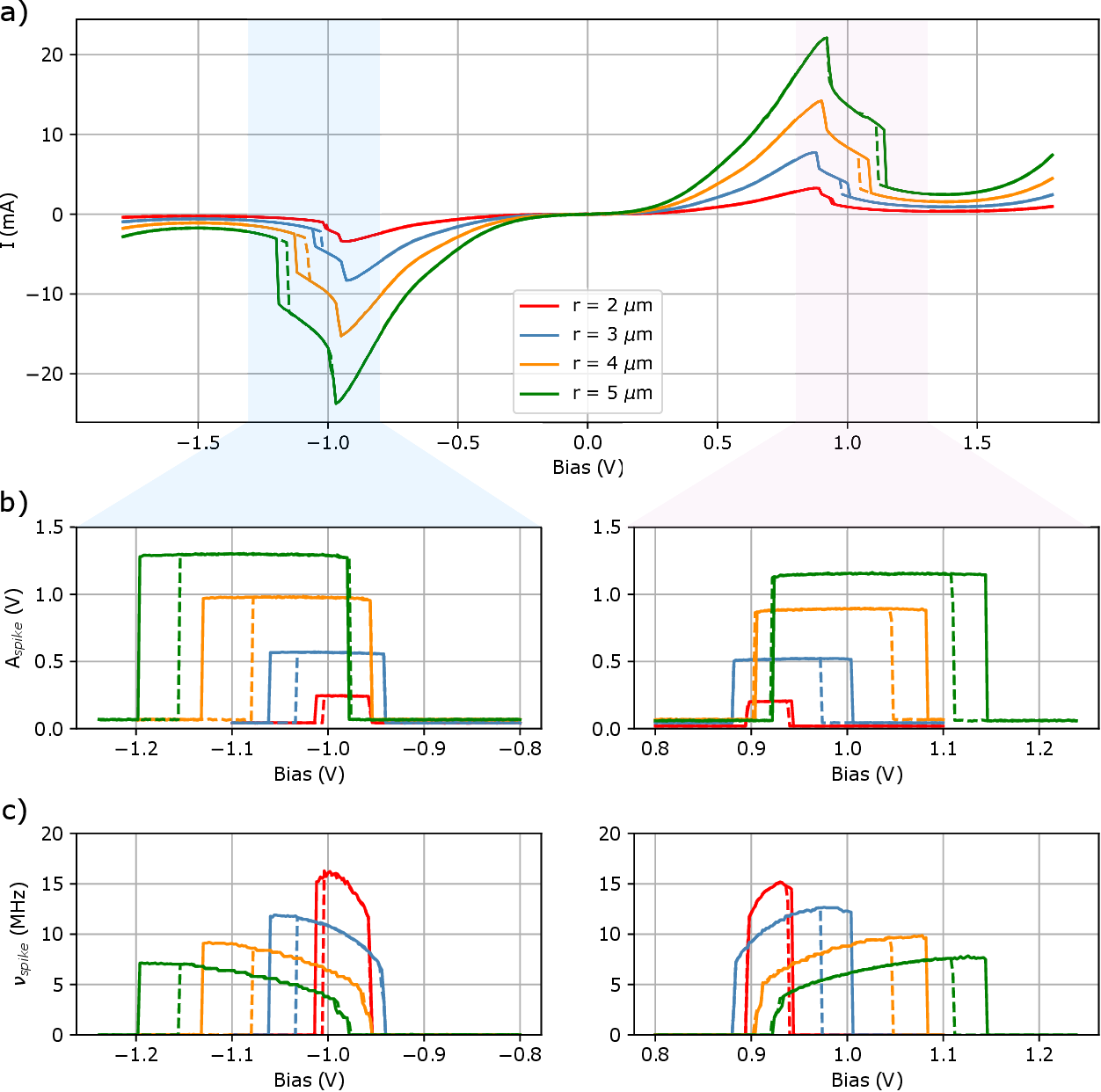}
\caption{a) Experimental I-V characteristics of RTDs having different circular mesa radii. b) Amplitude of the spike ($A_{spike}$) and c) firing rate ($\nu_{spike}$), exhibited by RTDs with different mesa-radius when biased across their correspondent NDR regions.}
\label{fig_many_radius}
\end{figure*}

Interestingly, Fig. \ref{fig_time_encoding} shows that the spike profile is influenced by the applied voltage. For $\delta V=-110$ mV (second row in Fig. \ref{fig_time_encoding}) the RTD responses resemble an amplified version of the input pulse. However, for bias closer to the peak ($\delta V=-80$ mV and $\delta V=-50$ mV, third and fourth rows in Fig. \ref{fig_time_encoding}, respectively), the same amplified input pulse is also followed by distinct discharge and recovery phases, lasting approximately 90 ns. The recovery period is linked to the relative refractory period \cite{hejda2023artificial} of the RTD neuron's circuit. In this period, similiarly to biological neurons, the system is less likely excitable from new input stimuli. Therefore, Fig. \ref{fig_time_encoding} unveils that simply tuning the bias allows switching the RTD operation from an amplifier to a neuron incorporating a refractory time. These are novel and fundamental details to account when designing and optimizing neuromorphic hardware relying on RTD neurons, strengthening once again the importance of considering the bias as a critical trainable parameter.  
As a final detail, it is worth noting that spikes generated through perturbations in the PDR share qualitative features with those produced when bias the device in its NDR and without perturbations (see Fig. \ref{fig_IV_3um}(c)). In both cases, the spike evolution is described by two fast and two slow phases, but when the device is biased in the PDR, an amplified profile of the input pulse emerges in between the first fast and subsequent slow phases. Moreover, while continuous spiking is achieved when the bias is set within the NDR, the PDR allows only a number of spikes limited by the single pulse duration.

\section{Influence of RTD design parameters}
\label{sec_design}

This section studies the influence of key system parameters, such as the RTD mesa radius and the overall circuit inductance and capacitance, in the spike encoding properties of RTD neurons. These insights provide valuable guidance for the future development of neuromorphic opto-electronic hardware architectures built with RTD neurons at their core. 

\subsection{RTD mesa radius}

Previous studies have shown that RTDs having circular mesa shape with larger radii allow more current to flow through the DBQW, resulting in shifted I-V characteristics \cite{figueiredo2001electric, romeira2023brain}. However, the specific effects of mesa radius on the spiking behaviour of RTDs remain unexplored. To address this, RTDs with mesa radii of 2 $\mu$m, 3 $\mu$m, 4 $\mu$m, and 5 $\mu$m were experimentally characterized in terms of their I-V curves, spike amplitudes, and firing rates. The results, summarized in Fig. \ref{fig_many_radius}, reveal key trends and implications for future optimised RTD designs.
The I-V characteristics of the RTDs are displayed in Fig. \ref{fig_many_radius}(a), showing a quasi-symmetric behaviour around the origin since the layerstack and overall fabricated RTD structure were targeted as symmetric as possible.
RTDs with larger mesa radii exhibit a shift in their I-V characteristics toward higher currents, consistent with the increased flow of electrons through the DBQW \cite{figueiredo2001electric, romeira2023brain}. Additionally, larger mesa radii expand the NDR region widening the spiking region. 
The corresponding spike amplitude ($A_{spike}$) and firing rate ($\nu_{spike}$) when biasing the RTDs across their NDR regions are shown in Fig. \ref{fig_many_radius}(b) and in Fig. \ref{fig_many_radius}(c), respectively. Figure \ref{fig_many_radius}(b, left panel) shows a zoom of the NDR regions under reverse bias, while Fig. \ref{fig_many_radius}(b, right panel) those under forward bias operation. Spike amplitudes and firing rates are zero outside the NDR regions but become non-zero within these regions, confirming the presence of spiking activity.
Notably, RTDs designed with larger radii fire electrical spikes having a larger amplitude, as the limit cycle orbit extends across the wider NDR region shown in Fig. \ref{fig_many_radius}(a) (see section \ref{sec_LC}).
\begin{figure}[t]
\includegraphics[width=0.4\textwidth]{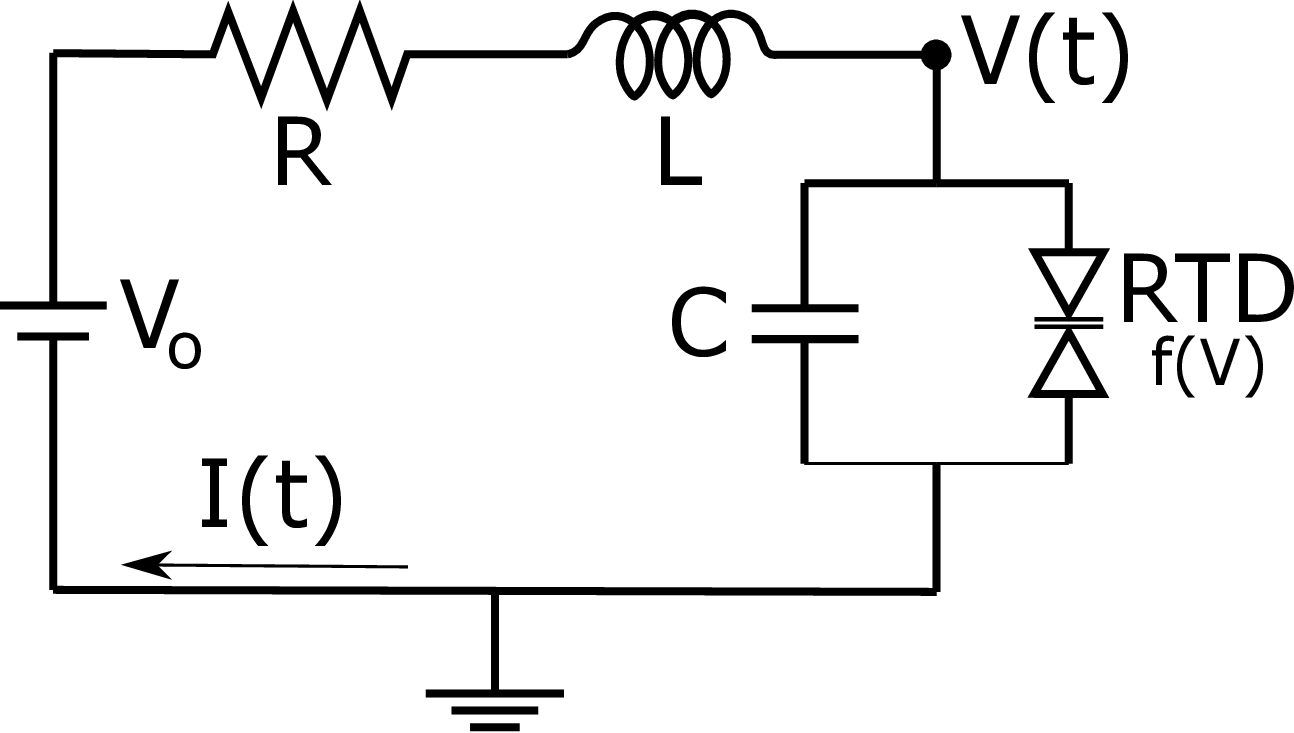}
\caption{RTD circuit scheme, described by an I-V function $f(V)$ coupled to circuit parameters R, L, and C. The inductance $L=L_{rtd}+L_{ext}$ and capacitance $C=C_{rtd}+C_{ext}$ account for RTD intrinsic ($L_{rtd}$ and $C_{rtd}$) and external ($L_{ext}$ and $C_{ext}$, the latter in paraller to $C_{rtd}$) components that expand the tunability of the RTD neuron circuit.}
\label{fig_circuit}
\end{figure}
The spike amplitude remains constant across the NDR region, independently on the RTD mesa radius, and is more pronounced under reverse bias configurations. 
Conversely, larger mesa radii reduce the firing rate (Fig. \ref{fig_many_radius}(c)), demonstrating an inverse relationship between spike amplitude and firing rate. 
Notably, Fig. \ref{fig_many_radius}(c) also highlights that RTDs with larger radii exhibit more restricted firing rate ranges spread over broader NDR regions. These devices are better suited for detecting external stimuli and maintaining consistent firing rates, even in noisy environments, due to their robustness against input voltage fluctuations. In contrast, RTDs with smaller radii offer wider firing rate ranges and hence encoding possibilities but are more susceptible to noise, making them ideal for tasks requiring stimulus distinction based on firing rate.

These findings have significant implications for neural network design. RTD neurons with smaller radii can specialize in tasks requiring diverse firing rates, while larger-radius RTDs are advantageous in noisy environments or when signal amplification is necessary. Networks can incorporate sub-networks tailored to specific functions, using RTDs with varying mesa radii. For example, spikes traveling through noisy network regions can be amplified by cascading RTDs with increasing radii, calibrated to generate higher-amplitude spikes as needed. This modular approach enhances the network’s robustness and adaptability for complex computational tasks.

\subsection{RTD circuit inductance and capacitance}
\label{sec_LC}

\begin{figure}[t]
\includegraphics[width=0.4\textwidth]{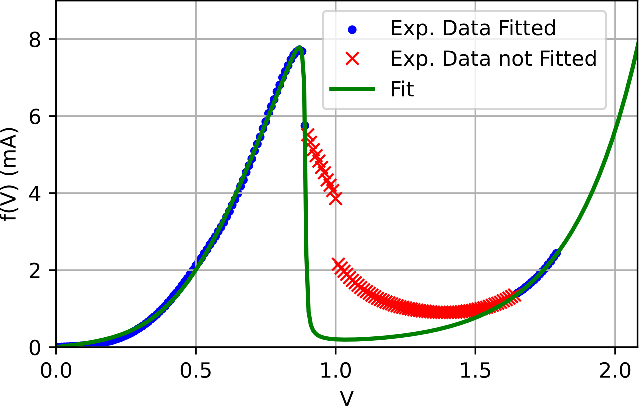}
\caption{Fit of f(V) to the experimental I-V characteristic of the RTD with 3 $\mu$m radius. Simulation parameters: $a=8.18\times 10^{-4}$ A, $b=8.535\times 10^{-2}$ V, $c=18.432\times 10^{-2}$ V, $d=7.209\times 10^{-4}$ V, $n_1=0.2066$, $n_2=0.1055$, $h=1.6025\times 10^{-6}$ A.}
\label{fig_fit}
\end{figure}

\begin{figure*}[t]
\includegraphics[width=1\textwidth]{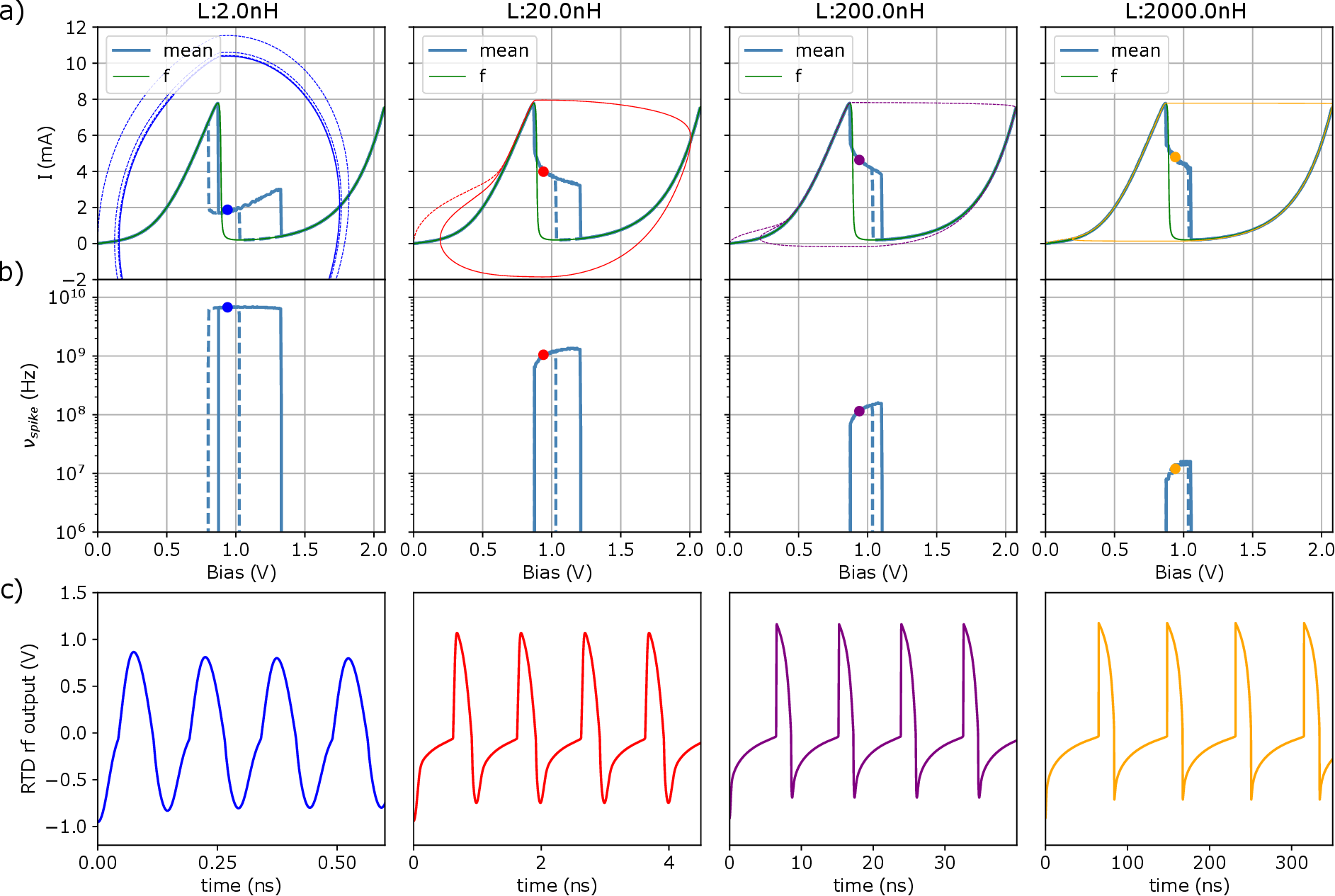}
\caption{Simulations of an RTD neuron circuit with six different inductance values $L$, indicated at the top of each column, and capacitance fixed at $C=250\times 10^{-15}$ F and resistance at $R=0.1$ $\Omega$. In panel (a), the I-V characteristics were evaluated by averaging the output current $I(t)$ (blue curve) and comparing it to the RTD’s pure quantum mechanical description $f$ across all bias values $V_0$. A highlighted coloured orbit illustrates the evolution of $V(t)$ and $I(t)$ during a spike, for $V_0=0.94$ V. Panel (b) shows the dependency of the firing rate on $L$, while panel (c) depicts the spiking evolution $V(t)$ over time for a bias of $V_0=0.94$ V, a value belonging to the NDR region for all six $L$ scenarios. Note that different timescales are used in the x-axis. }
\label{fig_L}
\end{figure*}

\begin{figure*}[t]
\includegraphics[width=1\textwidth]{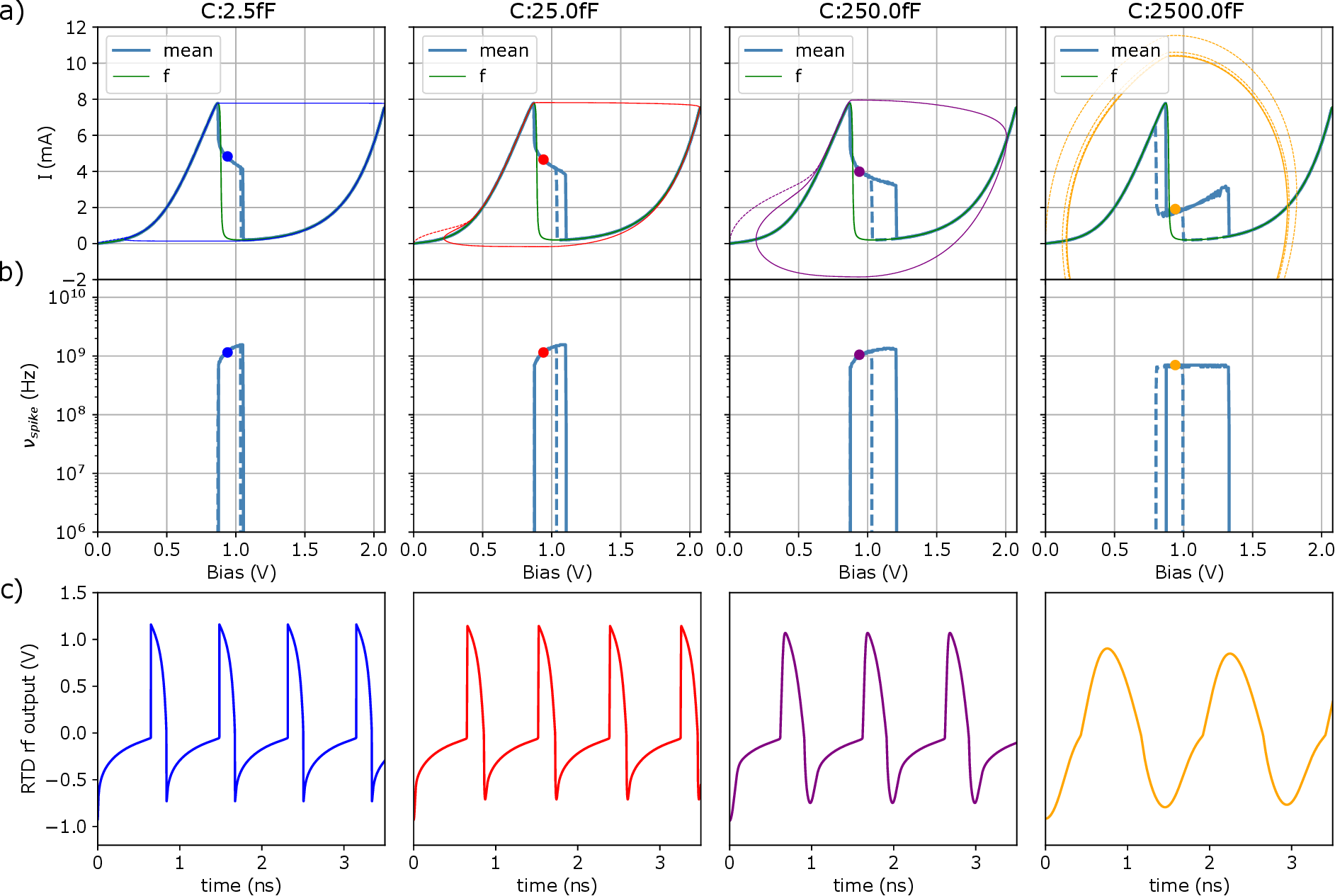}
\caption{Simulations of an RTD neuron circuit with five different capacitance values $C$, indicated at the top of each column, and inductance fixed at $L=20\times 10^{-9}$ H and resistance at $R=0.1$ $\Omega$. In panel (a), the I-V characteristics were evaluated by averaging the output current $I(t)$ (blue curve) and comparing it to the RTD’s pure quantum mechanical description $f$ across all bias values $V_0$. A highlighted coloured orbit illustrates the evolution of $V(t)$ and $I(t)$ during a spike, for $V_0=0.94$ V. Panel (b) shows the dependency of the firing rate on $C$, while panel (c) depicts the spiking evolution $V(t)$ over time for a bias of $V_0=0.94$ V, a value belonging to the NDR region of all six $C$ scenarios. }
\label{fig_C}
\end{figure*}

This section explores how to enhance the spiking rate of RTD neurons measured experimentally by tuning key circuit parameters. This analysis is performed numerically in order to have access to variables not directly accessible in the experiments, such as rapid current variations during the RTD spiking dynamics, and provide a deeper understanding of the system. The RTD circuit scheme we refer to is shown in Fig. \ref{fig_circuit}. By applying Kirchhoff's laws to the circuit, one obtains:
\begin{align}
    \label{eq:V}
    C\frac{dV(t)}{dt} &= I-f(V(t))\\
    \label{eq:I}
    L\frac{dI(t)}{dt} &= V_0-V(t)-RI(t),
\end{align}
\noindent In Eqs. \ref{eq:V} and \ref{eq:I}, $V(t)$ and $I(t)$ indicate the voltage and the total current across the RTD and capacitor C, while $R$ and $L$ indicate the circuit resistance and inductance, respectively. The parameters $L=L_{rtd}+L_{ext}$ and $C=C_{rtd}+C_{ext}$ account for RTD intrinsic ($L_{rtd}$ and $C_{rtd}$) and external ($L_{ext}$ and $C_{ext}$) components, with the latter aiming to extend the tunability of the RTD neuron circuit.
The function $f(V)$ appearing in Eq. \ref{eq:V} describes the RTD I-V characteristics with an expression derived directly from the quantum model in \cite{schulman1996physics}:
\begin{equation}
\label{eq:f}
    \begin{aligned}
        f(V) = &a\ln\left[\frac{1+e^{(q/k_BT)(b-c+n_1V)}}{1+e^{(q/k_BT)(b-c-n_1V)}}\right]\\
        &\times\left[\frac{\pi}{2}+tan^{-1}\left(\frac{c-n_1V}{d}\right)\right]+h\left[e^{(q/k_BT)n_2V}-1\right],
    \end{aligned}
\end{equation} 
where $T$ is the temperature, $q_e$ the electron charge and $k_B$ the Boltzmann's constant, while other parameters $a$, $b$, $c$, $d$, $n_1$, $n_2$ and $h$ depend on the RTD epilayer properties (e.g., geometry and doping levels) \cite{schulman1996physics}. 
These are chosen to fit the experimental I-V characteristic of the RTD having 3 $\mu$m mesa radius. The fit, shown as a green curve in Fig. \ref{fig_fit}, accounts for all experimental data (blue dots) and in particular for the sharp NDR transition just after the peak. The fit does not consider NDR points related to spiking activity (and some valley points labeled by red crosses in Fig. \ref{fig_fit}). 
Given f, the I-V characteristics accounting for the circuit parameters can be extracted by integrating Eqs. \ref{eq:V}-\ref{eq:I} for different dc bias $V_o$ and averaging over the correspondent simulated output time-trace $I(t)$, using this current value in the I-V. Note that this procedure resembles the average value of $I(t)$ measured experimentally for each bias by the current reader connected to the dc port of the bias tee (see Fig. \ref{fig_exp_scheme}). 
Figure \ref{fig_L}(a-b) shows the simulated I-V characteristics and correspondent spiking rate maps, respectively, for four inductance values ranging over four orders of magnitude, from $L=2\times 10^{-9}$ H to $L=2000\times 10^{-9}$ H (corresponding to the four columns in Fig. \ref{fig_L}). 
Each I-V curve in Fig. \ref{fig_L}(a) is measured by repeating the integration of Eqs. \ref{eq:V}-\ref{eq:I} for dc bias $V_o$ (x-axis in Fig. \ref{fig_L}(a)) from 0 V to 2.08 V (light blue continuous line) and viceversa (light blue dashed line). 
Figure \ref{fig_L}(a) also shows in green line the fit function $f(V)$ (Eq. \ref{eq:f}) used in the simulations, for reference. The same function $f$ is used in all simulations.

We first note that all four inductance scenarios studied allow spiking regimes ($\nu_{spike}\neq 0$ in Fig. \ref{fig_L}(b)) when the RTD is biased within the NDR region of the correspondent I-V curve (Fig. \ref{fig_L}(a)), and in agreement with the experimental findings in Fig. \ref{fig_IV_3um}. Notably, the NDR regions are also where the effects of circuit parameters are more pronounced, leading to deviations from $f(V)$. Remarkably, the NDR region excluded from the experimental fit of f(V) in Fig. \ref{fig_fit} naturally emerges when coupling f(V) to the circuit through Eqs. \ref{eq:I} and \ref{eq:f}. This suggests that the RTD spiking dynamics in the NDR region is an excitable process relying on resonance tunnelling effects within the DBQW (assumed instantaneous in the model), with timescales linked to the RTD neuron circuit parameters.
Increasing the inductance by a factor of 10 reduces the firing rate range by a factor of 10, from $\nu_{spike} \approx 10$ GHz for $L=2$ nH, down to $\nu_{spike}\approx 10$ MHz for $L=2$ $\mu$H, as shown in Fig. \ref{fig_L}(b)). 
The reduction in firing rate with $L$ is also shown by the temporal spiking responses plotted in Fig. \ref{fig_L}(c), under a 0.94 V bias (a value that belongs to the NDR regions of all four inductance scenarios investigated - see pairs of coloured dots in Fig. \ref{fig_L}(a-b)).
To grasp the physical origins behind the lower firing speed for larger inductance in the RTD circuit, the starting point is to map the (voltage) spiking temporal traces (for example the ones in Fig. \ref{fig_L}(c), obtained under a 0.94 V bias) in their correspondent limit cycle in the I-V plane (see the coloured orbits in Fig. \ref{fig_L}(a)), and observe that not only the voltage $V(t)$, but also the current $I(t)$ across the RTD are subject to variations during the spike evolution.
While for the lowest $L=2$ nH case, the orbit is more rounded (Fig. \ref{fig_L}(a) - first column) and described by a smooth oscillation over time (Fig. \ref{fig_L}(c) - first column), for larger $L$ the orbit becomes stiffer with emerging slow-fast temporal spike profiles, as predicted by the lower stiffness coefficient $\mu = \sqrt{C/L}$ obtained for larger $L$ \cite{ortega2021bursting}.  
By observing the limit cycle trajectories linked to a spike profile (see Fig. \ref{fig_L}(a) - $L\geq 20$ nH), four stages can be recognized: two stages where $V(t)$ experiences a larger variation while leaving $I(t)$ almost constant, and other two stages where both $V(t)$ and $I(t)$ vary steadily \cite{ortega2021bursting}. The inductance opposes to current variations in the form of an opposite electromotive force ($V_{0,L}=-L dI/dt$) and affects therefore the two stages of the spike evolution with wider current variation. On the other hand, the fast voltage phases of the spike evolution are controlled by the capacitance C of the RTD neuron circuit. We studied this dependence by performing simulations where C was varied over four orders of magnitude, from 2.5 fF to 2500 fF, while L was left fixed at $L=20$ nH. We assume that is the external component $C_{ext}$ of $C$ varying, while the intrinsic component $C_{rtd}$, linked to the RTD mesa radius, is the same. In this way all I-V characteristics show the same peak current. The results, shown in Fig. \ref{fig_C}, describe smoother spike fast phases while increasing C, with a maximum spike rate reduction of less than a factor 10 ($<<10^5$ obtained by varying L in Fig. \ref{fig_L}(b) by five order of magnitudes). Since the first fast phase of the spike evolution links to the integration time of the RTD neuron \cite{robertson2025ultrafast}, the capacitance becomes a parameter that allows slightly to chance this neuron property. Critically, as the stiffness coefficient $\mu = \sqrt{C/L}$ increases, either by decreasing L (as in Fig. \ref{fig_L}(a), left columns) and speeding up the slow current phases, or increasing C (as in Fig. \ref{fig_C}(a), right columns) and slowing down the fast voltage phase, the RTD spike gradually loses its slow-fast phases, flattening the corresponding firing rate variation within the NDR region (Fig. \ref{fig_L}(b), Fig. \ref{fig_C}(b), respectively). Consequently, this removes any rate encoding capability of the RTD neuron circuit. 

These results underscore the importance of carefully calibrating the inductance, capacitance and stiffness coefficient when designing RTD-based neuromorphic systems to achieve the desired spike operational speed and firing rate variability within the NDR region. Moreover, these findings are complementary to the earlier discussions on the impact of the RTD mesa radius. Specifically, the observed firing rate variations across RTDs having diverse mesa radii (see Fig. \ref{fig_many_radius}(c)) align with the numerical capacitance study. Since the RTD intrinsic capacitance increases with the mesa radius, it slightly impacts the firing rate. Additionally, a higher capacitance corresponds to a higher stiffness coefficient, which our simulations indicate it progressively constrains the firing rate variation within the NDR region. This explains the reduced firing rate variability in RTDs with larger mesa radii. 
Moreover, both inductance and capacitance affect the width of the bistable region at the end of the NDR, providing tuning capabilities for spiking memory effects \cite{donati2024spiking}. 
Together, these insights provide a deeper understanding of how circuit and device-level parameters influence RTD spiking dynamics.
Notably, the inductance study reported in Fig. \ref{fig_L} shows that for $L=2$ $\mu$H the RTD I-V curve, spike firing rate, and the widths of the NDR and bistability regions, closely match the experimental data in Fig. \ref{fig_IV_3um}. However, such an inductance would require an impractical large on-chip transmission line. We therefore attribute this inductance to the instrumentation connected to our RTD. Specifically, the bias-tee in Fig. \ref{fig_exp_scheme}, which has a bandwidth of 10 MHz – 12 GHz and an estimated internal inductance of a few $\mu H$, appears to be the limiting factor for the experimentally observed spiking rate. 
Consequently, fitting f(V) to the experimental NDR region (red crosses in Fig. \ref{fig_fit}) would inherently account for these experimental circuitry effects, even before coupling to the circuit parameters via Eqs. \ref{eq:V} and \ref{eq:I}, leading to deviations between numerical and experimental results.

\section{Rate Encoding Example}
\label{sec_app}

\begin{figure}[ht]
    \centering
    \includegraphics[width=1\linewidth]{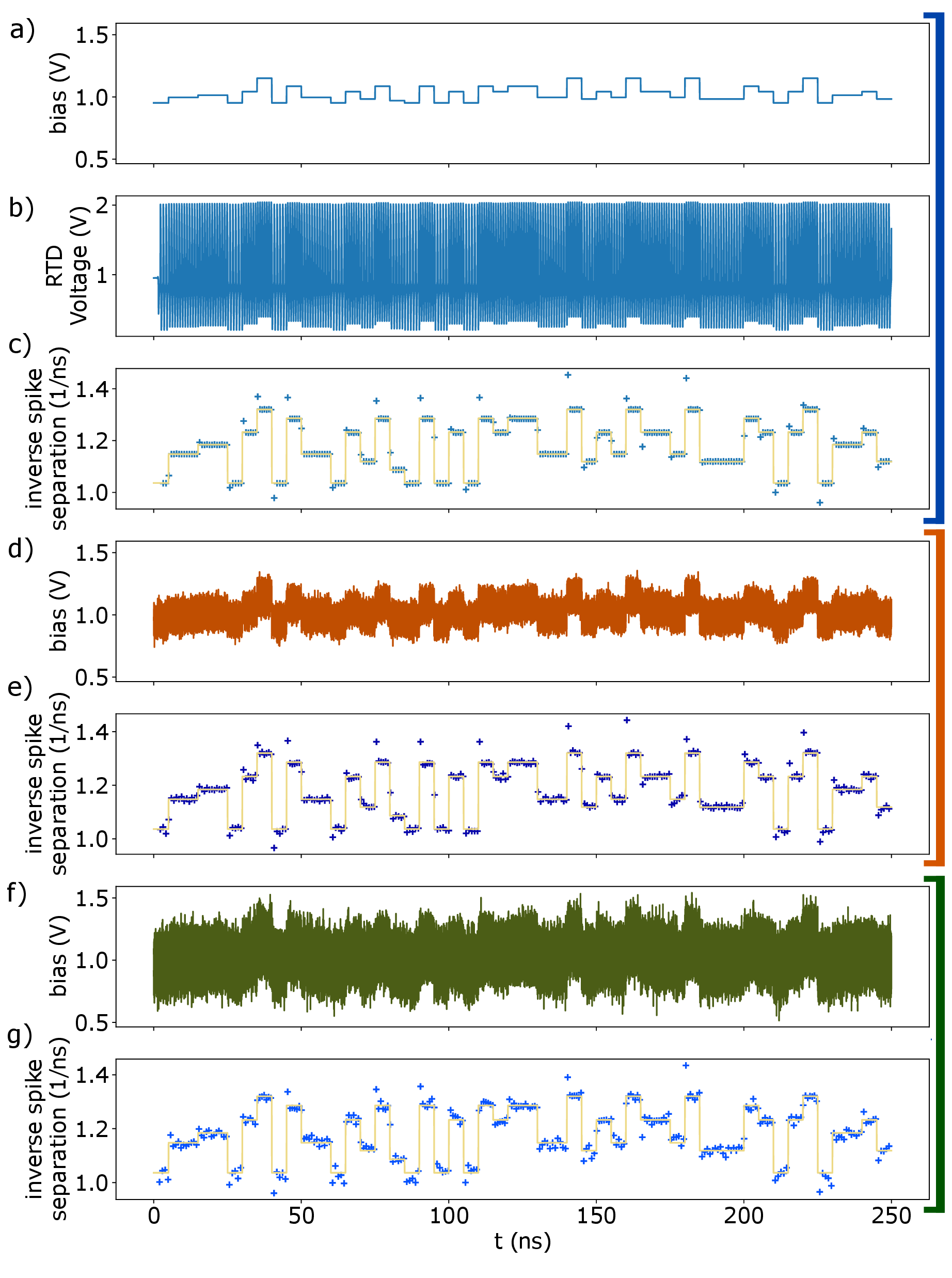}
    \caption{Simulation of a spike rate encoding of an 8 level word. a) Modulation of the bias voltage to the RTD. Each symbol was held for 5ns. b) Voltage across the RTD, showing the constant spiking at different rates, according to the bias modulation. c) Reconstruction of the signal from spike interval. At the time of occurrence of each spike, the reciprocal of the time since the previous spike is plotted, to approximate the inverse of the spike rate, recovering the modulation signal. d,f) Original modulation signal with normally distributed noise (standard deviation $\sigma=0.05$, $\sigma=0.1$) added. e,g) Reconstruction of signal by inverse spike interval, showing a good reconstruction of the non-noisy signal. Gold lines in c,e,g show the target function to recover.}
    \label{fig:word}
\end{figure}

Figure \ref{fig:word} shows an example of the RTD spike rate encoding scheme applied to a randomly generated word of eight-level symbols. Each level is mapped to a bias voltage within the NDR of the RTD neuron circuit, so that each level elicits a different spiking rate. This demonstration is conducted numerically because the bandwidth limitation of the bias tee in the current setup restricts input pulse durations to approximately 20 ns before filtering effects become significant. This duration is insufficient to produce multiple spikes and establish a spike rate. Instead, we consider an RTD with circuit parameters of $L=20$ nH and $C=250$ fF, as characterized in Fig. \ref{fig_C}, yielding a spike rate range between $0.65$ GHz and $1.35$ GHz.

The bias voltage across the RTD was varied between eight voltage levels within the NDR between 0.95V and 1.15V (spaced such that the resulting spike rates at each level were equally spaced), holding each level for 5 ns to allow several spikes. This modulation is shown in Fig. \ref{fig:word} (a). Fig. \ref{fig:word} (b) shows the response of the RTD with spikes occurring at rates according to the applied bias voltage. The instantaneous spiking rate was approximated from the voltage trace by the inverse of the inter-spike interval calculated at each spike (i.e. the difference between the time of a spike and the time of the preceding spike). Fig. \ref{fig:word} (c) shows the approximate spike rate at the moment of each spike, recreating the five-level word input to the bias modulation. Additionally, the spike rate encoding was performed on a noisy version of the original modulation. Gaussian noise with zero mean and a standard deviation of $\sigma=0.05$ (Fig. \ref{fig:word} (d)) and $\sigma=1$ (Fig. \ref{fig:word} (f)) were added to the voltage modulation. Since the frequency of the noise is much higher than the spiking frequency, the RTD spike rate encoding effectively filters out the noise and can make a good reconstruction of the signal from the inverse spike interval (Fig. \ref{fig:word} (e)(g)). This suggests the ability to effectively encode and decode noisy signals at high bit rate using spiking RTD neurons.

\section{Conclusion}
\label{sec_conclusions}
This work investigates experimentally and in theory the neuromorphic properties of RTD spiking neurons, emphasizing their potential for spike-based encoding mechanisms. The spike firing rate of an RTD can be tuned by adjusting the bias across the NDR region of its I-V characteristic, providing a basis for spike firing rate encoding. Additionally, we investigated encoding information using the latency time of spike firing in response to input stimuli. The study reveals a highly selective mechanism where only input pulses within a narrow amplitude range, on the order of a few mV above the spiking threshold, result in differentiated firing delays. This selectivity can be further controlled by fine-tuning the RTD bias, potentially enabling time-to-first-spike protocols in faster, optimized RTD neuron circuits. However, for bias below but far from the peak of the I-V curve, we also find that the elicited spikes show a weaker recovery phase (independently of the input pulse amplitude), which makes it easier to bypass the refractory period of the device.

The influence of RTD mesa radius and circuit inductance on spiking dynamics was also analyzed. RTDs with smaller radii are more suitable for firing rate encoding, as they offer a broader firing rate range. In contrast, larger-radius RTDs, which have a narrower firing rate range spread across a wider NDR region, are better suited for operation in noisy environments due to their robustness against voltage fluctuations. These results link to the higher capacitance of RTDs with larger mesa radius, which is numerically shown to slightly control the fast voltage phase time of the spike evolution. On the other hand, the inductance in the circuit resists the wide current variations during the spike evolution and significantly impacts the RTD's spike response by limiting the firing rate.
In this way, the operational speed is mainly controlled by the inductance of the circuit, with possibilities to be fine-tuned by the capacitance. 
Both the inductance and capacitance also tune the bistability width at the NDR end, which is critical for spiking flip-flop memory effects \cite{donati2024spiking}.
Moreover, the stiffness coefficient also needs proper calibration, as higher stiffness coefficient gradually smooth the slow-fast dynamics during the spike evolution, removing the firing rate capability within the NDR region. These findings provide valuable guidance for designing RTD-based neuromorphic systems, potentially co-working with external photonic hardware such as photo-detectors \cite{donati2024spiking}, offering exciting potential for photonic-electronic neuromorphic hardware.
\newline

\paragraph*{acknowledgments} 
The authors acknowledge support from the UKRI Turing AI Acceleration Fellowships Programme (EP/V025198/1), from the EU Pathfinder Open project ‘SpikePro’, from the Fraunhofer Centre for Applied Photonics, FCAP, and from the Fundação para a Ciência e a Tecnologia (FCT) project 2022.03392.PTDC–META-LED. 
\newline

\paragraph*{Data availability}
The data that support the findings of this article are openly available at \cite{Donati2025_Data}.
\newline

\paragraph*{Author Contributions}
G.D. performed the experimental and numerical study of RTDs; D.O. realized the numerical application of rate-encoding; J.F., V.C. and E.M. designed the RTDs, which where then fabricated at TU/e; J.R. and X.P. helped setting up the lab experiment; J.F. and B.R. supported the numerical analysis; A.H. supervised the work. All authors contributed to the revision of the manuscript.

\nocite{*}

\bibliography{apssamp}

\begin{thebibliography}{32}%
\makeatletter
\providecommand \@ifxundefined [1]{%
 \@ifx{#1\undefined}
}%
\providecommand \@ifnum [1]{%
 \ifnum #1\expandafter \@firstoftwo
 \else \expandafter \@secondoftwo
 \fi
}%
\providecommand \@ifx [1]{%
 \ifx #1\expandafter \@firstoftwo
 \else \expandafter \@secondoftwo
 \fi
}%
\providecommand \natexlab [1]{#1}%
\providecommand \enquote  [1]{``#1''}%
\providecommand \bibnamefont  [1]{#1}%
\providecommand \bibfnamefont [1]{#1}%
\providecommand \citenamefont [1]{#1}%
\providecommand \href@noop [0]{\@secondoftwo}%
\providecommand \href [0]{\begingroup \@sanitize@url \@href}%
\providecommand \@href[1]{\@@startlink{#1}\@@href}%
\providecommand \@@href[1]{\endgroup#1\@@endlink}%
\providecommand \@sanitize@url [0]{\catcode `\\12\catcode `\$12\catcode `\&12\catcode `\#12\catcode `\^12\catcode `\_12\catcode `\%12\relax}%
\providecommand \@@startlink[1]{}%
\providecommand \@@endlink[0]{}%
\providecommand \url  [0]{\begingroup\@sanitize@url \@url }%
\providecommand \@url [1]{\endgroup\@href {#1}{\urlprefix }}%
\providecommand \urlprefix  [0]{URL }%
\providecommand \Eprint [0]{\href }%
\providecommand \doibase [0]{https://doi.org/}%
\providecommand \selectlanguage [0]{\@gobble}%
\providecommand \bibinfo  [0]{\@secondoftwo}%
\providecommand \bibfield  [0]{\@secondoftwo}%
\providecommand \translation [1]{[#1]}%
\providecommand \BibitemOpen [0]{}%
\providecommand \bibitemStop [0]{}%
\providecommand \bibitemNoStop [0]{.\EOS\space}%
\providecommand \EOS [0]{\spacefactor3000\relax}%
\providecommand \BibitemShut  [1]{\csname bibitem#1\endcsname}%
\let\auto@bib@innerbib\@empty
\bibitem [{\citenamefont {Schuman}\ \emph {et~al.}(2022)\citenamefont {Schuman}, \citenamefont {Kulkarni}, \citenamefont {Parsa}, \citenamefont {Mitchell}, \citenamefont {Date},\ and\ \citenamefont {Kay}}]{schuman2022opportunities}%
  \BibitemOpen
  \bibfield  {author} {\bibinfo {author} {\bibfnamefont {C.~D.}\ \bibnamefont {Schuman}}, \bibinfo {author} {\bibfnamefont {S.~R.}\ \bibnamefont {Kulkarni}}, \bibinfo {author} {\bibfnamefont {M.}~\bibnamefont {Parsa}}, \bibinfo {author} {\bibfnamefont {J.~P.}\ \bibnamefont {Mitchell}}, \bibinfo {author} {\bibfnamefont {P.}~\bibnamefont {Date}},\ and\ \bibinfo {author} {\bibfnamefont {B.}~\bibnamefont {Kay}},\ }\bibfield  {title} {\bibinfo {title} {Opportunities for neuromorphic computing algorithms and applications},\ }\href@noop {} {\bibfield  {journal} {\bibinfo  {journal} {Nature Computational Science}\ }\textbf {\bibinfo {volume} {2}},\ \bibinfo {pages} {10} (\bibinfo {year} {2022})}\BibitemShut {NoStop}%
\bibitem [{\citenamefont {Eshraghian}\ \emph {et~al.}(2023)\citenamefont {Eshraghian}, \citenamefont {Ward}, \citenamefont {Neftci}, \citenamefont {Wang}, \citenamefont {Lenz}, \citenamefont {Dwivedi}, \citenamefont {Bennamoun}, \citenamefont {Jeong},\ and\ \citenamefont {Lu}}]{eshraghian2023training}%
  \BibitemOpen
  \bibfield  {author} {\bibinfo {author} {\bibfnamefont {J.~K.}\ \bibnamefont {Eshraghian}}, \bibinfo {author} {\bibfnamefont {M.}~\bibnamefont {Ward}}, \bibinfo {author} {\bibfnamefont {E.~O.}\ \bibnamefont {Neftci}}, \bibinfo {author} {\bibfnamefont {X.}~\bibnamefont {Wang}}, \bibinfo {author} {\bibfnamefont {G.}~\bibnamefont {Lenz}}, \bibinfo {author} {\bibfnamefont {G.}~\bibnamefont {Dwivedi}}, \bibinfo {author} {\bibfnamefont {M.}~\bibnamefont {Bennamoun}}, \bibinfo {author} {\bibfnamefont {D.~S.}\ \bibnamefont {Jeong}},\ and\ \bibinfo {author} {\bibfnamefont {W.~D.}\ \bibnamefont {Lu}},\ }\bibfield  {title} {\bibinfo {title} {Training spiking neural networks using lessons from deep learning},\ }\href@noop {} {\bibfield  {journal} {\bibinfo  {journal} {Proceedings of the IEEE}\ } (\bibinfo {year} {2023})}\BibitemShut {NoStop}%
\bibitem [{\citenamefont {Waldrop}(2016)}]{waldrop2016chips}%
  \BibitemOpen
  \bibfield  {author} {\bibinfo {author} {\bibfnamefont {M.~M.}\ \bibnamefont {Waldrop}},\ }\bibfield  {title} {\bibinfo {title} {The chips are down for moore’s law},\ }\href@noop {} {\bibfield  {journal} {\bibinfo  {journal} {Nature News}\ }\textbf {\bibinfo {volume} {530}},\ \bibinfo {pages} {144} (\bibinfo {year} {2016})}\BibitemShut {NoStop}%
\bibitem [{\citenamefont {Furber}\ \emph {et~al.}(2014)\citenamefont {Furber}, \citenamefont {Galluppi}, \citenamefont {Temple},\ and\ \citenamefont {Plana}}]{furber2014spinnaker}%
  \BibitemOpen
  \bibfield  {author} {\bibinfo {author} {\bibfnamefont {S.~B.}\ \bibnamefont {Furber}}, \bibinfo {author} {\bibfnamefont {F.}~\bibnamefont {Galluppi}}, \bibinfo {author} {\bibfnamefont {S.}~\bibnamefont {Temple}},\ and\ \bibinfo {author} {\bibfnamefont {L.~A.}\ \bibnamefont {Plana}},\ }\bibfield  {title} {\bibinfo {title} {The spinnaker project},\ }\href@noop {} {\bibfield  {journal} {\bibinfo  {journal} {Proceedings of the IEEE}\ }\textbf {\bibinfo {volume} {102}},\ \bibinfo {pages} {652} (\bibinfo {year} {2014})}\BibitemShut {NoStop}%
\bibitem [{\citenamefont {Painkras}\ \emph {et~al.}(2013)\citenamefont {Painkras}, \citenamefont {Plana}, \citenamefont {Garside}, \citenamefont {Temple}, \citenamefont {Galluppi}, \citenamefont {Patterson}, \citenamefont {Lester}, \citenamefont {Brown},\ and\ \citenamefont {Furber}}]{painkras2013spinnaker}%
  \BibitemOpen
  \bibfield  {author} {\bibinfo {author} {\bibfnamefont {E.}~\bibnamefont {Painkras}}, \bibinfo {author} {\bibfnamefont {L.~A.}\ \bibnamefont {Plana}}, \bibinfo {author} {\bibfnamefont {J.}~\bibnamefont {Garside}}, \bibinfo {author} {\bibfnamefont {S.}~\bibnamefont {Temple}}, \bibinfo {author} {\bibfnamefont {F.}~\bibnamefont {Galluppi}}, \bibinfo {author} {\bibfnamefont {C.}~\bibnamefont {Patterson}}, \bibinfo {author} {\bibfnamefont {D.~R.}\ \bibnamefont {Lester}}, \bibinfo {author} {\bibfnamefont {A.~D.}\ \bibnamefont {Brown}},\ and\ \bibinfo {author} {\bibfnamefont {S.~B.}\ \bibnamefont {Furber}},\ }\bibfield  {title} {\bibinfo {title} {Spinnaker: A 1-w 18-core system-on-chip for massively-parallel neural network simulation},\ }\href@noop {} {\bibfield  {journal} {\bibinfo  {journal} {IEEE Journal of Solid-State Circuits}\ }\textbf {\bibinfo {volume} {48}},\ \bibinfo {pages} {1943} (\bibinfo {year} {2013})}\BibitemShut {NoStop}%
\bibitem [{\citenamefont {DeBole}\ \emph {et~al.}(2019)\citenamefont {DeBole}, \citenamefont {Taba}, \citenamefont {Amir}, \citenamefont {Akopyan}, \citenamefont {Andreopoulos}, \citenamefont {Risk}, \citenamefont {Kusnitz}, \citenamefont {Otero}, \citenamefont {Nayak}, \citenamefont {Appuswamy} \emph {et~al.}}]{debole2019truenorth}%
  \BibitemOpen
  \bibfield  {author} {\bibinfo {author} {\bibfnamefont {M.~V.}\ \bibnamefont {DeBole}}, \bibinfo {author} {\bibfnamefont {B.}~\bibnamefont {Taba}}, \bibinfo {author} {\bibfnamefont {A.}~\bibnamefont {Amir}}, \bibinfo {author} {\bibfnamefont {F.}~\bibnamefont {Akopyan}}, \bibinfo {author} {\bibfnamefont {A.}~\bibnamefont {Andreopoulos}}, \bibinfo {author} {\bibfnamefont {W.~P.}\ \bibnamefont {Risk}}, \bibinfo {author} {\bibfnamefont {J.}~\bibnamefont {Kusnitz}}, \bibinfo {author} {\bibfnamefont {C.~O.}\ \bibnamefont {Otero}}, \bibinfo {author} {\bibfnamefont {T.~K.}\ \bibnamefont {Nayak}}, \bibinfo {author} {\bibfnamefont {R.}~\bibnamefont {Appuswamy}}, \emph {et~al.},\ }\bibfield  {title} {\bibinfo {title} {Truenorth: Accelerating from zero to 64 million neurons in 10 years},\ }\href@noop {} {\bibfield  {journal} {\bibinfo  {journal} {Computer}\ }\textbf {\bibinfo {volume} {52}},\ \bibinfo {pages} {20} (\bibinfo {year} {2019})}\BibitemShut {NoStop}%
\bibitem [{\citenamefont {Davies}\ \emph {et~al.}(2018)\citenamefont {Davies}, \citenamefont {Srinivasa}, \citenamefont {Lin}, \citenamefont {Chinya}, \citenamefont {Cao}, \citenamefont {Choday}, \citenamefont {Dimou}, \citenamefont {Joshi}, \citenamefont {Imam}, \citenamefont {Jain} \emph {et~al.}}]{davies2018loihi}%
  \BibitemOpen
  \bibfield  {author} {\bibinfo {author} {\bibfnamefont {M.}~\bibnamefont {Davies}}, \bibinfo {author} {\bibfnamefont {N.}~\bibnamefont {Srinivasa}}, \bibinfo {author} {\bibfnamefont {T.-H.}\ \bibnamefont {Lin}}, \bibinfo {author} {\bibfnamefont {G.}~\bibnamefont {Chinya}}, \bibinfo {author} {\bibfnamefont {Y.}~\bibnamefont {Cao}}, \bibinfo {author} {\bibfnamefont {S.~H.}\ \bibnamefont {Choday}}, \bibinfo {author} {\bibfnamefont {G.}~\bibnamefont {Dimou}}, \bibinfo {author} {\bibfnamefont {P.}~\bibnamefont {Joshi}}, \bibinfo {author} {\bibfnamefont {N.}~\bibnamefont {Imam}}, \bibinfo {author} {\bibfnamefont {S.}~\bibnamefont {Jain}}, \emph {et~al.},\ }\bibfield  {title} {\bibinfo {title} {Loihi: A neuromorphic manycore processor with on-chip learning},\ }\href@noop {} {\bibfield  {journal} {\bibinfo  {journal} {Ieee Micro}\ }\textbf {\bibinfo {volume} {38}},\ \bibinfo {pages} {82} (\bibinfo {year} {2018})}\BibitemShut {NoStop}%
\bibitem [{\citenamefont {Schemmel}\ \emph {et~al.}(2017)\citenamefont {Schemmel}, \citenamefont {Kriener}, \citenamefont {M{\"u}ller},\ and\ \citenamefont {Meier}}]{schemmel2017accelerated}%
  \BibitemOpen
  \bibfield  {author} {\bibinfo {author} {\bibfnamefont {J.}~\bibnamefont {Schemmel}}, \bibinfo {author} {\bibfnamefont {L.}~\bibnamefont {Kriener}}, \bibinfo {author} {\bibfnamefont {P.}~\bibnamefont {M{\"u}ller}},\ and\ \bibinfo {author} {\bibfnamefont {K.}~\bibnamefont {Meier}},\ }\bibfield  {title} {\bibinfo {title} {An accelerated analog neuromorphic hardware system emulating nmda-and calcium-based non-linear dendrites},\ }in\ \href@noop {} {\emph {\bibinfo {booktitle} {2017 International Joint Conference on Neural Networks (IJCNN)}}}\ (\bibinfo {organization} {IEEE},\ \bibinfo {year} {2017})\ pp.\ \bibinfo {pages} {2217--2226}\BibitemShut {NoStop}%
\bibitem [{\citenamefont {Ashtiani}\ \emph {et~al.}(2022)\citenamefont {Ashtiani}, \citenamefont {Geers},\ and\ \citenamefont {Aflatouni}}]{ashtiani2022chip}%
  \BibitemOpen
  \bibfield  {author} {\bibinfo {author} {\bibfnamefont {F.}~\bibnamefont {Ashtiani}}, \bibinfo {author} {\bibfnamefont {A.~J.}\ \bibnamefont {Geers}},\ and\ \bibinfo {author} {\bibfnamefont {F.}~\bibnamefont {Aflatouni}},\ }\bibfield  {title} {\bibinfo {title} {An on-chip photonic deep neural network for image classification},\ }\href@noop {} {\bibfield  {journal} {\bibinfo  {journal} {Nature}\ }\textbf {\bibinfo {volume} {606}},\ \bibinfo {pages} {501} (\bibinfo {year} {2022})}\BibitemShut {NoStop}%
\bibitem [{\citenamefont {Biasi}\ \emph {et~al.}(2023)\citenamefont {Biasi}, \citenamefont {Franchi}, \citenamefont {Cerini},\ and\ \citenamefont {Pavesi}}]{biasi2023array}%
  \BibitemOpen
  \bibfield  {author} {\bibinfo {author} {\bibfnamefont {S.}~\bibnamefont {Biasi}}, \bibinfo {author} {\bibfnamefont {R.}~\bibnamefont {Franchi}}, \bibinfo {author} {\bibfnamefont {L.}~\bibnamefont {Cerini}},\ and\ \bibinfo {author} {\bibfnamefont {L.}~\bibnamefont {Pavesi}},\ }\bibfield  {title} {\bibinfo {title} {An array of microresonators as a photonic extreme learning machine},\ }\href@noop {} {\bibfield  {journal} {\bibinfo  {journal} {APL Photonics}\ }\textbf {\bibinfo {volume} {8}} (\bibinfo {year} {2023})}\BibitemShut {NoStop}%
\bibitem [{\citenamefont {Feldmann}\ \emph {et~al.}(2019)\citenamefont {Feldmann}, \citenamefont {Youngblood}, \citenamefont {Wright}, \citenamefont {Bhaskaran},\ and\ \citenamefont {Pernice}}]{feldmann2019all}%
  \BibitemOpen
  \bibfield  {author} {\bibinfo {author} {\bibfnamefont {J.}~\bibnamefont {Feldmann}}, \bibinfo {author} {\bibfnamefont {N.}~\bibnamefont {Youngblood}}, \bibinfo {author} {\bibfnamefont {C.~D.}\ \bibnamefont {Wright}}, \bibinfo {author} {\bibfnamefont {H.}~\bibnamefont {Bhaskaran}},\ and\ \bibinfo {author} {\bibfnamefont {W.~H.}\ \bibnamefont {Pernice}},\ }\bibfield  {title} {\bibinfo {title} {All-optical spiking neurosynaptic networks with self-learning capabilities},\ }\href@noop {} {\bibfield  {journal} {\bibinfo  {journal} {Nature}\ }\textbf {\bibinfo {volume} {569}},\ \bibinfo {pages} {208} (\bibinfo {year} {2019})}\BibitemShut {NoStop}%
\bibitem [{\citenamefont {Lin}\ \emph {et~al.}(2018)\citenamefont {Lin}, \citenamefont {Rivenson}, \citenamefont {Yardimci}, \citenamefont {Veli}, \citenamefont {Luo}, \citenamefont {Jarrahi},\ and\ \citenamefont {Ozcan}}]{lin2018all}%
  \BibitemOpen
  \bibfield  {author} {\bibinfo {author} {\bibfnamefont {X.}~\bibnamefont {Lin}}, \bibinfo {author} {\bibfnamefont {Y.}~\bibnamefont {Rivenson}}, \bibinfo {author} {\bibfnamefont {N.~T.}\ \bibnamefont {Yardimci}}, \bibinfo {author} {\bibfnamefont {M.}~\bibnamefont {Veli}}, \bibinfo {author} {\bibfnamefont {Y.}~\bibnamefont {Luo}}, \bibinfo {author} {\bibfnamefont {M.}~\bibnamefont {Jarrahi}},\ and\ \bibinfo {author} {\bibfnamefont {A.}~\bibnamefont {Ozcan}},\ }\bibfield  {title} {\bibinfo {title} {All-optical machine learning using diffractive deep neural networks},\ }\href@noop {} {\bibfield  {journal} {\bibinfo  {journal} {Science}\ }\textbf {\bibinfo {volume} {361}},\ \bibinfo {pages} {1004} (\bibinfo {year} {2018})}\BibitemShut {NoStop}%
\bibitem [{\citenamefont {Ferry}(2020)}]{ferry2020transport}%
  \BibitemOpen
  \bibfield  {author} {\bibinfo {author} {\bibfnamefont {D.~K.}\ \bibnamefont {Ferry}},\ }\href@noop {} {\emph {\bibinfo {title} {Transport in Semiconductor Mesoscopic Devices}}}\ (\bibinfo  {publisher} {IOP Publishing},\ \bibinfo {year} {2020})\BibitemShut {NoStop}%
\bibitem [{\citenamefont {Ironside}\ \emph {et~al.}(2023)\citenamefont {Ironside}, \citenamefont {Romeira},\ and\ \citenamefont {Figueiredo}}]{ironside2023resonant}%
  \BibitemOpen
  \bibfield  {author} {\bibinfo {author} {\bibfnamefont {C.}~\bibnamefont {Ironside}}, \bibinfo {author} {\bibfnamefont {B.}~\bibnamefont {Romeira}},\ and\ \bibinfo {author} {\bibfnamefont {J.}~\bibnamefont {Figueiredo}},\ }\bibfield  {title} {\bibinfo {title} {Resonant tunnelling diode photonics devices and applications},\ }\href@noop {} {\bibfield  {journal} {\bibinfo  {journal} {IOP Publishing}\ } (\bibinfo {year} {2023})}\BibitemShut {NoStop}%
\bibitem [{\citenamefont {Cimbri}\ \emph {et~al.}(2022)\citenamefont {Cimbri}, \citenamefont {Wang}, \citenamefont {Al-Khalidi},\ and\ \citenamefont {Wasige}}]{cimbri2022resonant}%
  \BibitemOpen
  \bibfield  {author} {\bibinfo {author} {\bibfnamefont {D.}~\bibnamefont {Cimbri}}, \bibinfo {author} {\bibfnamefont {J.}~\bibnamefont {Wang}}, \bibinfo {author} {\bibfnamefont {A.}~\bibnamefont {Al-Khalidi}},\ and\ \bibinfo {author} {\bibfnamefont {E.}~\bibnamefont {Wasige}},\ }\bibfield  {title} {\bibinfo {title} {Resonant tunneling diodes high-speed terahertz wireless communications-a review},\ }\href@noop {} {\bibfield  {journal} {\bibinfo  {journal} {IEEE Transactions on Terahertz Science and Technology}\ }\textbf {\bibinfo {volume} {12}},\ \bibinfo {pages} {226} (\bibinfo {year} {2022})}\BibitemShut {NoStop}%
\bibitem [{\citenamefont {Nishida}\ \emph {et~al.}(2019)\citenamefont {Nishida}, \citenamefont {Nishigami}, \citenamefont {Diebold}, \citenamefont {Kim}, \citenamefont {Fujita},\ and\ \citenamefont {Nagatsuma}}]{nishida2019terahertz}%
  \BibitemOpen
  \bibfield  {author} {\bibinfo {author} {\bibfnamefont {Y.}~\bibnamefont {Nishida}}, \bibinfo {author} {\bibfnamefont {N.}~\bibnamefont {Nishigami}}, \bibinfo {author} {\bibfnamefont {S.}~\bibnamefont {Diebold}}, \bibinfo {author} {\bibfnamefont {J.}~\bibnamefont {Kim}}, \bibinfo {author} {\bibfnamefont {M.}~\bibnamefont {Fujita}},\ and\ \bibinfo {author} {\bibfnamefont {T.}~\bibnamefont {Nagatsuma}},\ }\bibfield  {title} {\bibinfo {title} {Terahertz coherent receiver using a single resonant tunnelling diode},\ }\href@noop {} {\bibfield  {journal} {\bibinfo  {journal} {Scientific reports}\ }\textbf {\bibinfo {volume} {9}},\ \bibinfo {pages} {18125} (\bibinfo {year} {2019})}\BibitemShut {NoStop}%
\bibitem [{\citenamefont {Romeira}\ \emph {et~al.}(2013)\citenamefont {Romeira}, \citenamefont {Javaloyes}, \citenamefont {Ironside}, \citenamefont {Figueiredo}, \citenamefont {Balle},\ and\ \citenamefont {Piro}}]{romeira2013excitability}%
  \BibitemOpen
  \bibfield  {author} {\bibinfo {author} {\bibfnamefont {B.}~\bibnamefont {Romeira}}, \bibinfo {author} {\bibfnamefont {J.}~\bibnamefont {Javaloyes}}, \bibinfo {author} {\bibfnamefont {C.~N.}\ \bibnamefont {Ironside}}, \bibinfo {author} {\bibfnamefont {J.~M.}\ \bibnamefont {Figueiredo}}, \bibinfo {author} {\bibfnamefont {S.}~\bibnamefont {Balle}},\ and\ \bibinfo {author} {\bibfnamefont {O.}~\bibnamefont {Piro}},\ }\bibfield  {title} {\bibinfo {title} {Excitability and optical pulse generation in semiconductor lasers driven by resonant tunneling diode photo-detectors},\ }\href@noop {} {\bibfield  {journal} {\bibinfo  {journal} {Optics express}\ }\textbf {\bibinfo {volume} {21}},\ \bibinfo {pages} {20931} (\bibinfo {year} {2013})}\BibitemShut {NoStop}%
\bibitem [{\citenamefont {Hejda}\ \emph {et~al.}(2022)\citenamefont {Hejda}, \citenamefont {Alanis}, \citenamefont {Ortega-Piwonka}, \citenamefont {Louren{\c{c}}o}, \citenamefont {Figueiredo}, \citenamefont {Javaloyes}, \citenamefont {Romeira},\ and\ \citenamefont {Hurtado}}]{hejda2022resonant}%
  \BibitemOpen
  \bibfield  {author} {\bibinfo {author} {\bibfnamefont {M.}~\bibnamefont {Hejda}}, \bibinfo {author} {\bibfnamefont {J.~A.}\ \bibnamefont {Alanis}}, \bibinfo {author} {\bibfnamefont {I.}~\bibnamefont {Ortega-Piwonka}}, \bibinfo {author} {\bibfnamefont {J.}~\bibnamefont {Louren{\c{c}}o}}, \bibinfo {author} {\bibfnamefont {J.}~\bibnamefont {Figueiredo}}, \bibinfo {author} {\bibfnamefont {J.}~\bibnamefont {Javaloyes}}, \bibinfo {author} {\bibfnamefont {B.}~\bibnamefont {Romeira}},\ and\ \bibinfo {author} {\bibfnamefont {A.}~\bibnamefont {Hurtado}},\ }\bibfield  {title} {\bibinfo {title} {Resonant tunneling diode nano-optoelectronic excitable nodes for neuromorphic spike-based information processing},\ }\href@noop {} {\bibfield  {journal} {\bibinfo  {journal} {Physical Review Applied}\ }\textbf {\bibinfo {volume} {17}},\ \bibinfo {pages} {024072} (\bibinfo {year} {2022})}\BibitemShut {NoStop}%
\bibitem [{\citenamefont {Hejda}\ \emph {et~al.}(2023)\citenamefont {Hejda}, \citenamefont {Malysheva}, \citenamefont {Owen-Newns}, \citenamefont {Ali Al-Taai}, \citenamefont {Zhang}, \citenamefont {Ortega-Piwonka}, \citenamefont {Javaloyes}, \citenamefont {Wasige}, \citenamefont {Dolores-Calzadilla}, \citenamefont {Figueiredo} \emph {et~al.}}]{hejda2023artificial}%
  \BibitemOpen
  \bibfield  {author} {\bibinfo {author} {\bibfnamefont {M.}~\bibnamefont {Hejda}}, \bibinfo {author} {\bibfnamefont {E.}~\bibnamefont {Malysheva}}, \bibinfo {author} {\bibfnamefont {D.}~\bibnamefont {Owen-Newns}}, \bibinfo {author} {\bibfnamefont {Q.~R.}\ \bibnamefont {Ali Al-Taai}}, \bibinfo {author} {\bibfnamefont {W.}~\bibnamefont {Zhang}}, \bibinfo {author} {\bibfnamefont {I.}~\bibnamefont {Ortega-Piwonka}}, \bibinfo {author} {\bibfnamefont {J.}~\bibnamefont {Javaloyes}}, \bibinfo {author} {\bibfnamefont {E.}~\bibnamefont {Wasige}}, \bibinfo {author} {\bibfnamefont {V.}~\bibnamefont {Dolores-Calzadilla}}, \bibinfo {author} {\bibfnamefont {J.~M.}\ \bibnamefont {Figueiredo}}, \emph {et~al.},\ }\bibfield  {title} {\bibinfo {title} {Artificial optoelectronic spiking neuron based on a resonant tunnelling diode coupled to a vertical cavity surface emitting laser},\ }\href@noop {} {\bibfield  {journal} {\bibinfo  {journal} {Nanophotonics}\ }\textbf {\bibinfo {volume} {12}},\ \bibinfo {pages} {857} (\bibinfo
  {year} {2023})}\BibitemShut {NoStop}%
\bibitem [{\citenamefont {Ortega-Piwonka}\ \emph {et~al.}(2021)\citenamefont {Ortega-Piwonka}, \citenamefont {Piro}, \citenamefont {Figueiredo}, \citenamefont {Romeira},\ and\ \citenamefont {Javaloyes}}]{ortega2021bursting}%
  \BibitemOpen
  \bibfield  {author} {\bibinfo {author} {\bibfnamefont {I.}~\bibnamefont {Ortega-Piwonka}}, \bibinfo {author} {\bibfnamefont {O.}~\bibnamefont {Piro}}, \bibinfo {author} {\bibfnamefont {J.}~\bibnamefont {Figueiredo}}, \bibinfo {author} {\bibfnamefont {B.}~\bibnamefont {Romeira}},\ and\ \bibinfo {author} {\bibfnamefont {J.}~\bibnamefont {Javaloyes}},\ }\bibfield  {title} {\bibinfo {title} {Bursting and excitability in neuromorphic resonant tunneling diodes},\ }\href@noop {} {\bibfield  {journal} {\bibinfo  {journal} {Physical Review Applied}\ }\textbf {\bibinfo {volume} {15}},\ \bibinfo {pages} {034017} (\bibinfo {year} {2021})}\BibitemShut {NoStop}%
\bibitem [{\citenamefont {Romeira}\ \emph {et~al.}(2014)\citenamefont {Romeira}, \citenamefont {Av{\'o}}, \citenamefont {Javaloyes}, \citenamefont {Balle}, \citenamefont {Ironside},\ and\ \citenamefont {Figueiredo}}]{romeira2014stochastic}%
  \BibitemOpen
  \bibfield  {author} {\bibinfo {author} {\bibfnamefont {B.}~\bibnamefont {Romeira}}, \bibinfo {author} {\bibfnamefont {R.}~\bibnamefont {Av{\'o}}}, \bibinfo {author} {\bibfnamefont {J.}~\bibnamefont {Javaloyes}}, \bibinfo {author} {\bibfnamefont {S.}~\bibnamefont {Balle}}, \bibinfo {author} {\bibfnamefont {C.~N.}\ \bibnamefont {Ironside}},\ and\ \bibinfo {author} {\bibfnamefont {J.~M.}\ \bibnamefont {Figueiredo}},\ }\bibfield  {title} {\bibinfo {title} {Stochastic induced dynamics in neuromorphic optoelectronic oscillators},\ }\href@noop {} {\bibfield  {journal} {\bibinfo  {journal} {Optical and Quantum Electronics}\ }\textbf {\bibinfo {volume} {46}},\ \bibinfo {pages} {1391} (\bibinfo {year} {2014})}\BibitemShut {NoStop}%
\bibitem [{\citenamefont {Donati}\ \emph {et~al.}(2024)\citenamefont {Donati}, \citenamefont {Owen-Newns}, \citenamefont {Robertson}, \citenamefont {Malysheva}, \citenamefont {Adair}, \citenamefont {Figueiredo}, \citenamefont {Romeira}, \citenamefont {Dolores-Calzadilla},\ and\ \citenamefont {Hurtado}}]{donati2024spiking}%
  \BibitemOpen
  \bibfield  {author} {\bibinfo {author} {\bibfnamefont {G.}~\bibnamefont {Donati}}, \bibinfo {author} {\bibfnamefont {D.}~\bibnamefont {Owen-Newns}}, \bibinfo {author} {\bibfnamefont {J.}~\bibnamefont {Robertson}}, \bibinfo {author} {\bibfnamefont {E.}~\bibnamefont {Malysheva}}, \bibinfo {author} {\bibfnamefont {A.}~\bibnamefont {Adair}}, \bibinfo {author} {\bibfnamefont {J.}~\bibnamefont {Figueiredo}}, \bibinfo {author} {\bibfnamefont {B.}~\bibnamefont {Romeira}}, \bibinfo {author} {\bibfnamefont {V.}~\bibnamefont {Dolores-Calzadilla}},\ and\ \bibinfo {author} {\bibfnamefont {A.}~\bibnamefont {Hurtado}},\ }\bibfield  {title} {\bibinfo {title} {Spiking flip-flop memory in resonant tunnelling diode neurons},\ }\href@noop {} {\bibfield  {journal} {\bibinfo  {journal} {Physical Review Letters}\ }\textbf {\bibinfo {volume} {133}},\ \bibinfo {pages} {267301} (\bibinfo {year} {2024})}\BibitemShut {NoStop}%
\bibitem [{\citenamefont {Sano}\ \emph {et~al.}(2001)\citenamefont {Sano}, \citenamefont {Murata}, \citenamefont {Otsuji}, \citenamefont {Akeyoshi}, \citenamefont {Shimizu},\ and\ \citenamefont {Sano}}]{sano200180}%
  \BibitemOpen
  \bibfield  {author} {\bibinfo {author} {\bibfnamefont {K.}~\bibnamefont {Sano}}, \bibinfo {author} {\bibfnamefont {K.}~\bibnamefont {Murata}}, \bibinfo {author} {\bibfnamefont {T.}~\bibnamefont {Otsuji}}, \bibinfo {author} {\bibfnamefont {T.}~\bibnamefont {Akeyoshi}}, \bibinfo {author} {\bibfnamefont {N.}~\bibnamefont {Shimizu}},\ and\ \bibinfo {author} {\bibfnamefont {E.}~\bibnamefont {Sano}},\ }\bibfield  {title} {\bibinfo {title} {An 80-gb/s optoelectronic delayed flip-flop ic using resonant tunneling diodes and uni-traveling-carrier photodiode},\ }\href@noop {} {\bibfield  {journal} {\bibinfo  {journal} {IEEE Journal of Solid-State Circuits}\ }\textbf {\bibinfo {volume} {36}},\ \bibinfo {pages} {281} (\bibinfo {year} {2001})}\BibitemShut {NoStop}%
\bibitem [{\citenamefont {Romeira}\ \emph {et~al.}(2017)\citenamefont {Romeira}, \citenamefont {Figueiredo},\ and\ \citenamefont {Javaloyes}}]{romeira2017delay}%
  \BibitemOpen
  \bibfield  {author} {\bibinfo {author} {\bibfnamefont {B.}~\bibnamefont {Romeira}}, \bibinfo {author} {\bibfnamefont {J.~M.}\ \bibnamefont {Figueiredo}},\ and\ \bibinfo {author} {\bibfnamefont {J.}~\bibnamefont {Javaloyes}},\ }\bibfield  {title} {\bibinfo {title} {Delay dynamics of neuromorphic optoelectronic nanoscale resonators: Perspectives and applications},\ }\href@noop {} {\bibfield  {journal} {\bibinfo  {journal} {Chaos: An Interdisciplinary Journal of Nonlinear Science}\ }\textbf {\bibinfo {volume} {27}} (\bibinfo {year} {2017})}\BibitemShut {NoStop}%
\bibitem [{\citenamefont {Slight}\ and\ \citenamefont {Ironside}(2007)}]{slight2007investigation}%
  \BibitemOpen
  \bibfield  {author} {\bibinfo {author} {\bibfnamefont {T.~J.}\ \bibnamefont {Slight}}\ and\ \bibinfo {author} {\bibfnamefont {C.~N.}\ \bibnamefont {Ironside}},\ }\bibfield  {title} {\bibinfo {title} {Investigation into the integration of a resonant tunnelling diode and an optical communications laser: Model and experiment},\ }\href@noop {} {\bibfield  {journal} {\bibinfo  {journal} {IEEE journal of quantum electronics}\ }\textbf {\bibinfo {volume} {43}},\ \bibinfo {pages} {580} (\bibinfo {year} {2007})}\BibitemShut {NoStop}%
\bibitem [{\citenamefont {Richmond}\ \emph {et~al.}(1987)\citenamefont {Richmond}, \citenamefont {Optican}, \citenamefont {Podell},\ and\ \citenamefont {Spitzer}}]{richmond1987temporal}%
  \BibitemOpen
  \bibfield  {author} {\bibinfo {author} {\bibfnamefont {B.~J.}\ \bibnamefont {Richmond}}, \bibinfo {author} {\bibfnamefont {L.~M.}\ \bibnamefont {Optican}}, \bibinfo {author} {\bibfnamefont {M.}~\bibnamefont {Podell}},\ and\ \bibinfo {author} {\bibfnamefont {H.}~\bibnamefont {Spitzer}},\ }\bibfield  {title} {\bibinfo {title} {Temporal encoding of two-dimensional patterns by single units in primate inferior temporal cortex. i. response characteristics},\ }\href@noop {} {\bibfield  {journal} {\bibinfo  {journal} {Journal of neurophysiology}\ }\textbf {\bibinfo {volume} {57}},\ \bibinfo {pages} {132} (\bibinfo {year} {1987})}\BibitemShut {NoStop}%
\bibitem [{\citenamefont {Middlebrooks}\ \emph {et~al.}(1994)\citenamefont {Middlebrooks}, \citenamefont {Clock}, \citenamefont {Xu},\ and\ \citenamefont {Green}}]{middlebrooks1994panoramic}%
  \BibitemOpen
  \bibfield  {author} {\bibinfo {author} {\bibfnamefont {J.~C.}\ \bibnamefont {Middlebrooks}}, \bibinfo {author} {\bibfnamefont {A.~E.}\ \bibnamefont {Clock}}, \bibinfo {author} {\bibfnamefont {L.}~\bibnamefont {Xu}},\ and\ \bibinfo {author} {\bibfnamefont {D.~M.}\ \bibnamefont {Green}},\ }\bibfield  {title} {\bibinfo {title} {A panoramic code for sound location by cortical neurons},\ }\href@noop {} {\bibfield  {journal} {\bibinfo  {journal} {Science}\ }\textbf {\bibinfo {volume} {264}},\ \bibinfo {pages} {842} (\bibinfo {year} {1994})}\BibitemShut {NoStop}%
\bibitem [{\citenamefont {Figueiredo}\ \emph {et~al.}(2001)\citenamefont {Figueiredo}, \citenamefont {Ironside},\ and\ \citenamefont {Stanley}}]{figueiredo2001electric}%
  \BibitemOpen
  \bibfield  {author} {\bibinfo {author} {\bibfnamefont {J.~L.}\ \bibnamefont {Figueiredo}}, \bibinfo {author} {\bibfnamefont {C.~N.}\ \bibnamefont {Ironside}},\ and\ \bibinfo {author} {\bibfnamefont {C.~R.}\ \bibnamefont {Stanley}},\ }\bibfield  {title} {\bibinfo {title} {Electric field switching in a resonant tunneling diode electroabsorption modulator},\ }\href@noop {} {\bibfield  {journal} {\bibinfo  {journal} {IEEE Journal of Quantum Electronics}\ }\textbf {\bibinfo {volume} {37}},\ \bibinfo {pages} {1547} (\bibinfo {year} {2001})}\BibitemShut {NoStop}%
\bibitem [{\citenamefont {Romeira}\ \emph {et~al.}(2023)\citenamefont {Romeira}, \citenamefont {Ad{\~a}o}, \citenamefont {Nieder}, \citenamefont {Al-Taai}, \citenamefont {Zhang}, \citenamefont {Hadfield}, \citenamefont {Wasige}, \citenamefont {Hejda}, \citenamefont {Hurtado}, \citenamefont {Malysheva} \emph {et~al.}}]{romeira2023brain}%
  \BibitemOpen
  \bibfield  {author} {\bibinfo {author} {\bibfnamefont {B.}~\bibnamefont {Romeira}}, \bibinfo {author} {\bibfnamefont {R.~R.}\ \bibnamefont {Ad{\~a}o}}, \bibinfo {author} {\bibfnamefont {J.~B.}\ \bibnamefont {Nieder}}, \bibinfo {author} {\bibfnamefont {Q.}~\bibnamefont {Al-Taai}}, \bibinfo {author} {\bibfnamefont {W.}~\bibnamefont {Zhang}}, \bibinfo {author} {\bibfnamefont {R.~H.}\ \bibnamefont {Hadfield}}, \bibinfo {author} {\bibfnamefont {E.}~\bibnamefont {Wasige}}, \bibinfo {author} {\bibfnamefont {M.}~\bibnamefont {Hejda}}, \bibinfo {author} {\bibfnamefont {A.}~\bibnamefont {Hurtado}}, \bibinfo {author} {\bibfnamefont {E.}~\bibnamefont {Malysheva}}, \emph {et~al.},\ }\bibfield  {title} {\bibinfo {title} {Brain-inspired nanophotonic spike computing: challenges and prospects},\ }\href@noop {} {\bibfield  {journal} {\bibinfo  {journal} {Neuromorphic Computing and Engineering}\ }\textbf {\bibinfo {volume} {3}},\ \bibinfo {pages} {033001} (\bibinfo {year} {2023})}\BibitemShut {NoStop}%
\bibitem [{\citenamefont {Schulman}\ \emph {et~al.}(1996)\citenamefont {Schulman}, \citenamefont {De~Los~Santos},\ and\ \citenamefont {Chow}}]{schulman1996physics}%
  \BibitemOpen
  \bibfield  {author} {\bibinfo {author} {\bibfnamefont {J.}~\bibnamefont {Schulman}}, \bibinfo {author} {\bibfnamefont {H.}~\bibnamefont {De~Los~Santos}},\ and\ \bibinfo {author} {\bibfnamefont {D.}~\bibnamefont {Chow}},\ }\bibfield  {title} {\bibinfo {title} {Physics-based rtd current-voltage equation},\ }\href@noop {} {\bibfield  {journal} {\bibinfo  {journal} {IEEE Electron Device Letters}\ }\textbf {\bibinfo {volume} {17}},\ \bibinfo {pages} {220} (\bibinfo {year} {1996})}\BibitemShut {NoStop}%
\bibitem [{\citenamefont {Robertson}\ \emph {et~al.}(2025)\citenamefont {Robertson}, \citenamefont {Black}, \citenamefont {Donati}, \citenamefont {Al-Taai}, \citenamefont {Malysheva}, \citenamefont {Romeira}, \citenamefont {Figueiredo}, \citenamefont {Calzadilla}, \citenamefont {Wasige},\ and\ \citenamefont {Hurtado}}]{robertson2025ultrafast}%
  \BibitemOpen
  \bibfield  {author} {\bibinfo {author} {\bibfnamefont {J.}~\bibnamefont {Robertson}}, \bibinfo {author} {\bibfnamefont {D.}~\bibnamefont {Black}}, \bibinfo {author} {\bibfnamefont {G.}~\bibnamefont {Donati}}, \bibinfo {author} {\bibfnamefont {Q.~R.~A.}\ \bibnamefont {Al-Taai}}, \bibinfo {author} {\bibfnamefont {E.}~\bibnamefont {Malysheva}}, \bibinfo {author} {\bibfnamefont {B.}~\bibnamefont {Romeira}}, \bibinfo {author} {\bibfnamefont {J.}~\bibnamefont {Figueiredo}}, \bibinfo {author} {\bibfnamefont {V.~D.}\ \bibnamefont {Calzadilla}}, \bibinfo {author} {\bibfnamefont {E.}~\bibnamefont {Wasige}},\ and\ \bibinfo {author} {\bibfnamefont {A.}~\bibnamefont {Hurtado}},\ }\bibfield  {title} {\bibinfo {title} {Ultrafast and compact photonic-electronic leaky integrate-and-fire circuits based upon resonant tunnelling diodes},\ }\href@noop {} {\bibfield  {journal} {\bibinfo  {journal} {arXiv preprint arXiv:2501.17133}\ } (\bibinfo {year} {2025})}\BibitemShut {NoStop}%
\bibitem [{\citenamefont {Donati}(2025)}]{Donati2025_Data}%
  \BibitemOpen
  \bibfield  {author} {\bibinfo {author} {\bibfnamefont {G.}~\bibnamefont {Donati}},\ }\href@noop {} {\bibinfo {title} {Data for:"\uppercase{s}piking rate and latency encoding with resonant tunnelling diode neuron circuits and design influences"}} (\bibinfo {year} {2025}),\ \bibinfo {note} {10.15129/fa675685-5b03-4866-a532-c291411f6ffc}\BibitemShut {NoStop}%
\end{thebibliography}%

\end{document}